\documentclass{new-assignment}
\usepackage{caption}
\usepackage{subcaption}
\usepackage{lipsum}
\usepackage{listings}
\usepackage{tabularx}
\usepackage{}
\usepackage{float}

\newcommand*{\name}{Bridget McGibbon}

\newcommand*{\school}{\textbf{Eindhoven University of Technology}}
\newcommand*{\department}{Department of Applied Physics}
\newcommand*{\group}{Science and Technology of Nuclear Fusion}
\newcommand*{\internship}{Internship Report}
\newcommand*{\supervisorone}{Dr. Matthias Hoelzl, Max Planck Institute for Plasma Physics}
\newcommand*{\supervisortwo}{Prof. Dr. Guido Huijsmans, CEA, Eindhoven University of Technology}
\newcommand*{\supervisorthree}{Dr. Rohan Ramasamy, Max Planck Institute for Plasma Physics}

\usepackage[
backend=biber,
style=ieee,
sorting=none
]{biblatex}

\newcommand*{\assignment}{Simulations of poloidal flow stabilization of ballooning modes in a classical l=2 stellarator using JOREK}

\begin{document}

\maketitle

\newpage

\section*{Abstract}
Magnetohydrodynamics (MHD), combining fluid dynamics and Maxwell's equations, provides a useful means of analysing the dynamic evolution of plasmas and plasma instabilities. JOREK is a non-linear MHD code which solves these equations in the context of magnetic confinement fusion. Originally developed for tokamaks, JOREK has been extended to model stellarators. In this project, E$\times$B poloidal flows are implemented in a classical l=2 stellarator configuration, by imposing a simple radial electric potential profile via initial conditions. The influence of this sheared background flow velocity on pressure-driven modes is interrogated, demonstrating a stabilizing effect when the shearing rate is comparable to the growth rate. This effect is observed for multiple toroidal modes and at different viscosities, demonstrating that the stabilization occurs as a result of shear decorrelation. Oscillations of the linear growth rate are observed in cases with higher flow speeds; this phenomenon is hypothesized to be due to phase misalignment between the poloidally coupled modes contributing to the ballooning mode. Some indicators are provided to support this, however analysis of this phenomenon is ongoing.

\thispagestyle{empty} 

\newpage

\tableofcontents
\thispagestyle{empty} 

\newpage

\section{Introduction}
The purpose of this section is to give context for the project concerned with edge instabilities in magnetic confinement fusion plasmas. The general features, advantages, and disadvantages of stellarators in comparison with the tokamak concept will be explained, followed by an overview of pressure-driven modes and poloidal flows. Following this, the research questions will be stated.\\

The information given here is intended as a starting point; a reader not familiar with fusion may need to consult the cited sources in order to entirely understand the project at hand.

\subsection{Nuclear Fusion} \label{sec:stellarators}
The impacts of climate change and the necessity to move away from the use of fossil fuels are well known. Nuclear fusion is a proposed alternative energy source aimed at harnessing the energy released when hydrogen nuclei fuse 
into helium, a  process which naturally occurs in a similar form in stars. This would provide a low-carbon energy source which does not fluctuate based on environmental factors, as do many renewable energy sources \cite{azarpour_review_2013}. The fuel is (in theory) abundant, with the desired hydrogen isotopes either occurring naturally or able to be extracted as byproducts of nuclear reactors, and doesn't result in long-term radioactive waste products \cite{freidberg_plasma_2007}. \\

Unfortunately, nuclear fusion only occurs when fuel is confined at extremely high temperatures and densities, due to the electromagnetic interaction holding the nuclei apart otherwise. A common parameter for measuring confinement is the Lawson criterion, or the ``triple product" of density, temperature, and energy confinement time. The triple product is an extension of Lawson's 1955 criterion which was derived from the basic requirement that the power produced by the reactor should be higher than the energy used to heat and maintain the reaction \cite{j_d_lawson_criteria_1955}. To achieve a self-sustaining deuterium-tritium reaction, the triple product should be 

\begin{equation}
    nT\tau_E > 1.5 \times 10^{21} m^{-3} keV s , 
\end{equation}
where the energy confinement time $\tau_E$ is defined as the energy of the plasma divided by the power required to sustain the plasma \cite{wesson_tokamaks_2004}.
Confining a plasma at the required temperature and density for long enough to create a viable energy source is not straight-forward. Several approaches to increasing the triple product exist. Inertial confinement methods aim to increase the triple product by maximizing the density of a fuel target, at the cost of decreased energy confinement times \cite{zohuri_confinement_2017}. Magnetic confinement methods use magnetic fields to contain a heated plasma, and these methods have lower densities but longer energy confinement times. Magnetic confinement in a toroidal device is largely accepted as the most feasible means of achieving the required confinement \cite{freidberg_plasma_2007}. \\

\subsection{Magnetic Confinement}
\begin{figure}
    \centering
    \includegraphics[width=0.7\textwidth]{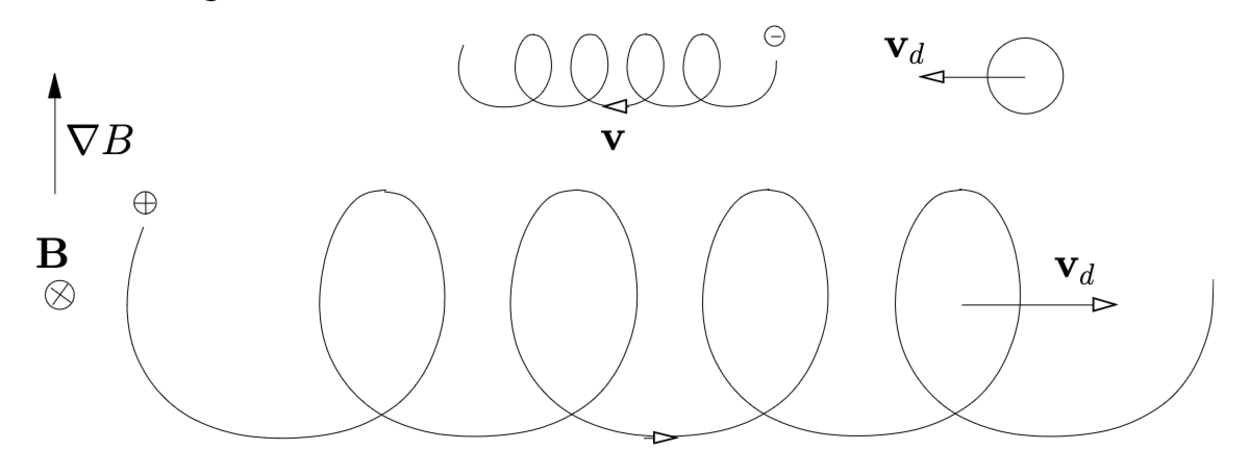}
    \caption{grad(B) drift of a negatively charged particle (top) and a positively charged particle (bottom). Note that the particles drift in the opposite direction. From \protect \cite{hutchinson_single-particle_nodate}.}
    \label{fig:gradB}
\end{figure}

First let us consider the general design of a magnetic confinement device. Consider confining a plasma in a cylinder, with a uniform magnetic field parallel to the axis around which the charged particles gyrate. It is clear that the ends of such a device would result in unacceptable losses. The cylinder can be bent into a torus, eliminating this concern. In doing so, we introduce a gradient to the magnetic field; the external current driving the magnetic field is necessarily more dense on the inner radius of the torus. Now, a positively charged particle gyrating in the field will have a smaller gyroradius in the region of higher magnetic field, and a larger gyroradius in the region of lower magnetic field, by equation \ref{eq:larmor}.
\begin{equation}\label{eq:larmor}
    r_L = \frac{mv_\perp}{qB}
\end{equation}
This causes particles to drift in a direction perpendicular to the magnetic field and its gradient, as shown in figure \ref{fig:gradB}. This effect is aptly named grad(B) drift. Positively and negatively charged particles will drift in the opposite direction, creating a vertical electric field.  \\

Let us now consider a single charged particle in uniform magnetic and electric fields, perpendicular to one another. As the particle gyrates in the magnetic field, the electric field accelerates it through half of its gyration and decelerates it through its other half. This causes the gyroradius to be larger on the side with higher electric potential, and causes the particle to drift in the direction normal to both the electric and magnetic fields. All particles, regardless of charge, experience the same drift, therefore a macroscopic plasma flow is created.  The velocity of this flow is
\begin{equation}
    \boldsymbol{v} = \frac{\boldsymbol{E} \times \boldsymbol{B}}{B^2}
        \label{eq:ecrossbdrift}
\end{equation}
In the case of the vertical electric field induced by the grad(B) drift (described above), the E$\times$B drift pushes the plasma to the outer edge of the torus, destroying confinement. \\

\begin{figure}[h]
    \centering
    \includegraphics[width=0.7\textwidth]{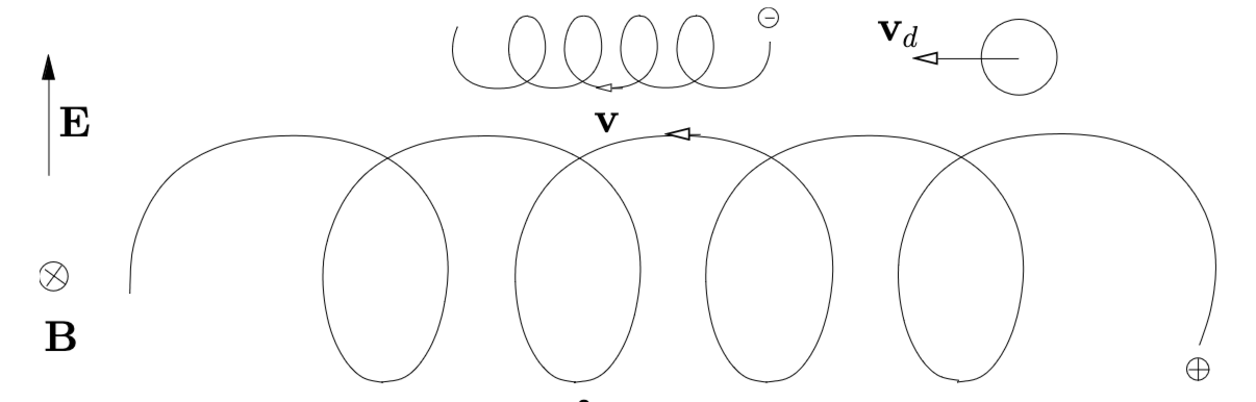}
    \caption{E$\times$B drift of a negatively charged particle (top) and a positively charged particle (bottom). Note that the particles drift in the same direction. From \protect \cite{hutchinson_single-particle_nodate}.}
    \label{fig:ExB}
\end{figure}

Luckily, this can be avoided if the particles are rotated from the top of the torus to the bottom at a sufficient rate such that the $\nabla B$ drift approximately cancels along the helical path and a global vertical electric field cannot form. This rotation is achieved by superimposing a poloidal field on top of the dominant toroidal field, creating a helical field.  The resulting helical field is sufficient to confine a plasma, excluding the effects of perturbations \cite{ongena_magnetic-confinement_2016}.  There are multiple ways to introduce this rotational transform. Two popular configurations are the tokamak, which achieves the rotation by driving a toroidal current inside the plasma, and the stellarator, which can also use toroidal current to create the rotation but is characterized by its use of external field coils to rotate the poloidal cross-section and/or twist the toroidal axis \cite{helander_stellarator_2012}. \\ 

\begin{figure}
    \label{fig:helicalfieldtokamak}
    \centering
    \includegraphics[width=0.5\textwidth]{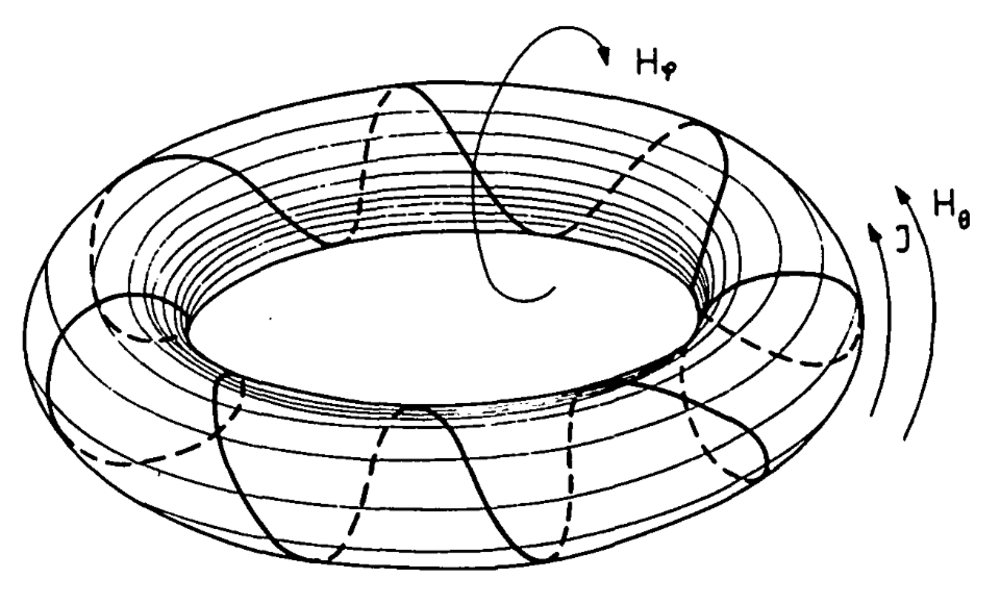}
    \caption{General magnetic configuration of a tokamak. The current J drives the toroidal magentic field H$_\theta$; The poloidal magentic field H$_\varphi$ is driven by external coils. From \protect\cite{artsimovich_tokamak_1972}.}
    \label{fig:tokamakfield}
\end{figure}

\subsection{Tokamaks}
Tokamaks are a heavily studied concept for magnetic confinement fusion. A tokamak achieves the required helical field via external coils which produce the toroidal component of the field. A solenoid in the center of the torus is then used to induce an electric current in the plasma, which in turn induces the poloidal part of the magnetic field \cite{wesson_tokamaks_2004}. Tokamaks experienced early success, with experiments performed in the 1960s demonstrating energy confinement times that were 50 times that of energy confinement times seen in most other devices at the time \cite{artsimovich_thermal_1967, braams_nuclear_2002}. Following further confirmation of these results, the fusion  community as a whole largely moved towards researching tokamaks, leaving research on other devices to lag behind \cite{lehnert_half_2013}. Tokamaks are also axisymmetric, so analytic and numerical studies can often be conducted in only 2 dimensions. This is a notable advantage over stellarators \cite{helander_stellarator_2012}.

\subsection{Stellarators}
Stellarators are another common type of magnetic confinement device which rely on external coils to generate the majority of the magnetic field \cite{xu_general_2016}, achieving the desired rotational transform primarily via torsion of the magnetic axis and/or toroidal variation of the poloidal cross-section. Because of this, stellarators are inherently non-axisymmetric, which increases the complexity of theoretical and computational analysis \cite{wagner_stellarators_1998}. However, advancing computational abilities allowed for stellarator designs to be optimized for better confinement from the late 1980s onwards \cite{nuhrenberg_quasi-helically_1988, grieger_physics_1992}. These improvements renewed interest in stellarators and their numerous advantages. While some stellarators are designed with significant toroidal net current, a notable advantage of many stellarators is that they can be designed to avoid net toroidal current altogether \cite{boozer_stellarator_2015}. In this way, they can avoid certain instabilities which pose significant issues for tokamaks \cite{weller_survey_2001}.  This is not to say that stellarators are completely stable, as all magnetic confinement configurations have potential for instabilities driven by steep pressure gradients.

\begin{figure}
    \centering
    \includegraphics[width=0.8\textwidth]{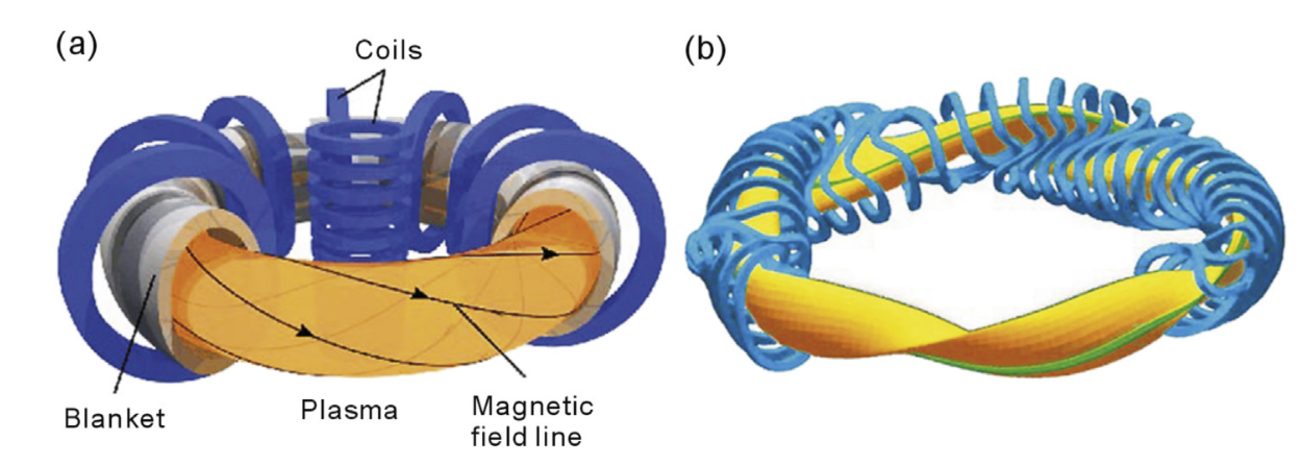}
    \caption{A tokamak (left) and a stellarator, W7-X (right). Note the toroidal variation of the plasma cross-section on the left. From \protect \cite{xu_general_2016}}
    \label{fig:tokvsstell}.
\end{figure}

.\\

\subsection{Plasma Stability }\label{sec:plasmastability}

The ideal magnetohydrodynamic\footnote{For more information on magnetohydrodynamics, refer to section \ref{sec:fulltoreducedMHD} or ref. \cite{freidberg_plasma_2007}} stability of a static magnetic confinement configuration can be represented as an energy functional as

\begin{equation}
    \delta W = \frac{1}{2}\int d^3 x (\frac{5}{3}p|\nabla \cdot \boldsymbol{\xi}|^2 + \frac{|\boldsymbol{B}|^2}{\mu_0}|\nabla \cdot \boldsymbol{\xi}+ 2\boldsymbol{\xi} \cdot \boldsymbol{\kappa}|^2 + \frac{|\boldsymbol{B}_{1\perp}|^2}{\mu_0} - J_\parallel(\boldsymbol{\xi}^*_\perp \times \boldsymbol{b} \cdot \boldsymbol{B}_{1\perp}) - 2 (\boldsymbol{\xi}_\perp \cdot \nabla p)(\boldsymbol{\xi}^*_\perp \cdot \boldsymbol{\kappa}))
    \label{eq:energyfunctional}
\end{equation} 

where W is potential energy, p is the plasma pressure, \textbf{B} is the equilibrium magnetic field, \textbf{b} is the unit vector parallel to the magnetic field, \textbf{J} is the current density, $\boldsymbol{\kappa} = \boldsymbol{b} \cdot \nabla \boldsymbol{b}$  is the curvature of the field, and {$\boldsymbol{\xi}$} is the perturbation from the equilibrium field \cite{freidberg_plasma_2007}. Subscripts $\parallel$, $\perp$, and 1 indicate directions parallel to the magnetic field, perpendicular to the magnetic field, and parallel to the perturbation, respectively. $\boldsymbol{\xi}^*$ is the conjugate of the perturbation. 
The energy function serves to describe the change in the potential energy due to a perturbation $\boldsymbol{\xi}$ of the equilibrium field \cite{greene_interchange_1968}. If all possible perturbations $\boldsymbol{\xi}$ result in a positive energy functional, then the configuration is stable. Otherwise the configuration has the potential to develop instabilities which can deteriorate or destroy confinement. \\

The first three terms of the energy functional are always stabilizing when present, representing the effects of plasma compression, field compression, and field line bending, respectively \cite{freidberg_plasma_2007}. The last two terms represent the destabilizing effects of current and the pressure gradient. The presence of the field line curvature $\boldsymbol{\kappa}$ within this last term should be noted; the pressure gradient term is only destabilizing in regions where the field line curvature is parallel to the pressure gradient. If the field line curvature opposes the pressure gradient, the resulting force perturbation is stabilizing. For this reason, areas where the curvature opposes the pressure gradient are referred to as areas of favourable curvature. \\

As stated in section \ref{sec:stellarators}, stellarators are often designed without net toroidal current in order to eliminate the current-driven instability term. This project concerns pressure-driven modes, which can be further categorized into interchange and ballooning modes.

\subsubsection{Interchange Modes} \label{sec:interchangemodes}
If the curvature of the magnetic field is in the same direction as the pressure gradient, it is possible for a perturbation to cause flux tubes of different pressures ``interchange" without causing field line bending. If the field line is perturbed outwards relative to the plasma, then the magnetic tension will decrease, while the opposite is true for perturbations inwards. Thus the mode is unstable. 

Because the interchange mode doesn't bend the field line, it affects the entire field line without much variation along the length. Thus, the field line may pass through areas of favourable and unfavourable curvature. The stability of the mode is then related to the average curvature \cite{freidberg_plasma_2007}.

\subsubsection{Ballooning Modes} \label{sec:ballooningmodes}
Ballooning instabilities also arise in areas of unfavourable curvature. However these modes are a perturbation that bends the field lines, becoming unstable when the pressure gradient exceeds the stabilizing effects of field line bending.
Since the amount of bending can vary along the field line, the ballooning mode is not constrained by good average curvature, and will still appear in regions of unfavourable curvature. In a tokamak, this occurs at the low-field side of the torus.\\

Pressure-driven modes limit the pressure that can be reached in the reactor and thereby its overall performance. It is thus desirable to introduce as much stabilization of these modes as possible. Poloidal flows represent one such stabilizing effect on pressure-driven modes, observed in both tokamaks and stellarators \cite{watanabe_poloidal_1992, burrell_effects_1997, burrell_role_2020}.

\subsection{Poloidal Flows} \label{subsec:poloidalflows}
Plasma flows can arise from several sources, with one of the most widely noted and studied sources being the E$\times$B drift (equation \ref{eq:ecrossbdrift}). Radial electric fields arise in both tokamaks and stellarators due to pressure gradient. This is described by the radial force balance

\begin{equation} \label{eq:radialforcebalance}
    E_r = \frac{\nabla P_i}{n_i Z_n e} - v_{\theta, i}B_\phi + v_{\phi,i}B_\theta , 
\end{equation}
where $E_r$ is the radial electric field, $P_i$ is the pressure as determined by the ions, $n_i$ is the ion number density, $Z_n e$ is the effective charge of the ions, $v_i$ is the velocity of the ions, and B is the magnetic field, with $\theta$ and $\phi$ being the poloidal and toroidal directions respectively. \\

Poloidal flows can stabilize a pressure-driven mode via several mechanisms. First of all, a poloidal velocity moves the plasma from the outboard to the inboard side of the plasma. Magnetic field lines are frozen into the plasma in the ideal model, and are thus convected with the plasma from areas of unfavourable curvature to areas of favourable curvature. Perturbations which only grow in areas of unfavourable curvature (i.e ballooning modes) are therefore stabilized by this effect. 
Poloidal flows are also known to affect the stability of pressure-driven modes by causing coupling from unstable to stable modes.
Additionally, sheared velocity tears apart the mode structures, also suppressing growth. \\

The equations determining the growth rates of the above-described modes are often linearized for ease of calculation. A given mode will generally grow at the linear rate until it saturates, when it becomes large enough to affect the background equilibrium. After the mode saturates, the more complicated non-linear dynamics must be considered \cite{wesson_tokamaks_2004}. In stellarators, it is often observed that modes which are linearly unstable are nonlinearly stable \cite{weller_significance_2006}. This is an advantage of stellarators, as many limits set by linear MHD are overly conservative, but it also presents an additional challenge in modelling \cite{zhou_approach_2021}.

\subsection{Purpose of the Project}
The stability of pressure-driven modes, as introduced in the above section, has serious implications for the performance of a reactor. Modes which grow uncontrollably can cause loss of confinement and damage to the reactor via sudden flux of heat and particles to the wall \cite{wilson_non-linear_nodate}. Even steady, saturated modes can still cause enhanced transport which lowers the efficiency of the reactor \cite{wesson_tokamaks_2004}.

JOREK is a simulation code for modelling non-linear MHD phenomena, containing many different physics models to adapt the code for different cases \cite{hoelzl_jorek_2021}. One adaptation is specific to stellarators and has been succesfully benchmarked against other simulation codes for multiple cases \cite{nikulsin_jorek3d_2022}. It doesn't yet include as many effects as are available in the code for tokamaks \cite{nikulsin_jorek3d_2022, hoelzl_jorek_2021}. One of the missing effects is the poloidal flow induced by a radial electric field. \\

This project concerns the implementation of poloidal flows into the stellarator model, with the goal of answering the following questions:
\begin{itemize}
    \item Can the stabilizing effect of flows on ballooning modes be captured in JOREK?

    \item Is the effect explained by the competition between mode growth and shearing rate? 

\end{itemize}

\newpage

\section{Theoretical Background}
This section will discuss how the MHD equations are implemented into JOREK. 

\subsection{Full to Reduced MHD} \label{sec:fulltoreducedMHD}

Magnetohydrodynamics (MHD) is a field of physics which combines fluid mechanics and Maxwell's equations to describe plasma as an electrically conducting fluid \cite{freidberg_plasma_2007}. The derivation of the single-fluid MHD model is not treated here, but can be performed from kinetic equations under the assumption that the plasma in question is highly collisional, and thus both ions and electrons have a Maxwellian distribution. This assumption does not hold completely for fusion plasmas, but the MHD model is sufficient for describing most large-scale phenomena nonetheless \cite{freidberg_plasma_2007, artsimovich_tokamak_1972, lehnert_half_2013}. \\ 

From either kinetic equations or from fluid dynamics and classical electrodynamics,  the following equations of full viscoresistive MHD can be derived:
\begin{subequations}
    \begin{equation}
            \frac{\delta \rho}{\delta t} + \nabla \cdot (\rho \boldsymbol{v}) = p 
    \end{equation}
    \begin{equation}
        \frac{\delta}{\delta t}(\rho \boldsymbol{v}) + \nabla \cdot (\rho \boldsymbol{v}\boldsymbol{v}) = \boldsymbol{j} \times \boldsymbol{B} - \nabla p + \rho \nu \Delta \boldsymbol{v}
    \end{equation}

    \begin{equation}
        \frac{\delta p}{\delta t}+ \boldsymbol{v} \cdot \nabla p + \gamma p \nabla\cdot v = (\gamma - 1)[\nabla\cdot(\kappa_\perp \nabla_\perp T + \kappa_\parallel \nabla_\parallel T + \frac{p}{\gamma - 1}\frac{D_\perp}{\rho}\nabla_\perp\rho)+S_e+\eta j^2]
    \end{equation}
    \begin{equation}
        \frac{\delta \boldsymbol{B}}{\delta t} = - \nabla \times \boldsymbol{E}
    \end{equation}
    \begin{equation}
        \nabla \times \boldsymbol{B} = \mu_0 \boldsymbol{j} 
    \end{equation}
    \begin{equation}
        \nabla \cdot \boldsymbol{B} = 0
    \end{equation}
    \begin{equation}
        \boldsymbol{E} = -\boldsymbol{v} \times \boldsymbol{B} + \eta \boldsymbol{j} 
    \end{equation}
    \begin{equation}
        p = \nabla \cdot (D_\perp \nabla_\perp \rho) + S_\rho
    \end{equation}
    \label{eq:MHDog}
\end{subequations}.

Throughout this report, $\rho$, $\boldsymbol{v}$, $p$, $\boldsymbol{j}$, $\boldsymbol{B}$, $\boldsymbol{E}$, and $T$ refer to the mass density, velocity, pressure, current density, magnetic field, electric field, and temperature, respectively. $\nu$ is the kinematic viscosity, $\gamma = \frac{5}{3}$ is the ratio of specific heats, and $\eta$ is the electrical resistivity. $\kappa$ and $D$ are the heat and mass diffusivities, and the $S_e$ and $S_p$ are heat and particle sources. The subscripts $\parallel$ and $\perp$ indicate directions parallel and perpendicular to the magnetic field. The gradient operators $\nabla_\parallel$ and $\nabla_\perp$ are defined as $\frac{\boldsymbol{B}}{B^2}\boldsymbol{B} \cdot \nabla$ and $\nabla-\nabla_\parallel$, respectively \cite{nikulsin_three-dimensional_2019}. \\

For stellarator applications, JOREK uses a reduced MHD model, meaning that the terms representing fast magnetosonic waves are eliminated\cite{nikulsin_jorek3d_2022}. This allows for coarser time resolution to be used, thus reducing the computational cost of simulation. Reduced MHD can be achieved by using either an ordering parameter which places a higher order on certain terms and eliminating them, or by using an ansatz which eliminates fast magnetosonic waves while keeping other higher order terms \cite{nikulsin_three-dimensional_2019}. JOREK uses the latter, allowing the model to retain more accurate energy conservation at the cost of increased equation complexity. \\

The total magnetic field is expressed in field-aligned coordinates as
\begin{equation}
    \boldsymbol{B} = \nabla\chi + \nabla\Psi \times \nabla\chi + \nabla\Omega \times \nabla\psi_v ,
\label{eq:Bansatz}
\end{equation}
where $\nabla \chi$ is the vacuum magnetic field, $\Psi$ and $\Omega$ are the $\chi$- and $\psi_v$- components of the vector potential of the induced magnetic field, respectively \cite{nikulsin_models_2021}. The terms on the right hand side are generally representative of the vacuum field, the bending of the field, and the compression of the field, respectively.

We also use the velocity ansatz
\begin{equation}
    \label{eq:velocityansatz}
    \boldsymbol{v} = \frac{\nabla\Phi \times \nabla\chi}{B_v^2} + v_\parallel \boldsymbol{B} + \nabla^\perp\zeta,
\end{equation}
with $B_v=|\nabla \chi|$ and $\nabla \Phi$ being the electric field. The terms on the right are generally representative of the E$\times$B flows, the field-aligned flows, and the compression of the fluid. The model used here will neglect the last two terms of ansatz \ref{eq:velocityansatz}, and the last term of ansatz \ref{eq:Bansatz}. \\

These ansatzes are applied to the MHD equations above to obtain equations for the evolution of electric potential $\Phi$, density $\rho$, pressure $p$, poloidal flux $\psi$, current density $j$, and vorticity $\omega$. The equations which are finally implemented into the code are 

\begin{subequations} \label{reducedMHD}
    \begin{equation}
        \nabla \cdot (\frac{\rho}{B_v^2}\nabla^\perp\frac{\delta\Phi}{\delta t}) = \frac{B_v}{2}[\frac{\rho}{B_v^2},\frac{(\Phi,\Phi)}{B_v^2}] 
        + B_v[\frac{\rho \tilde{\omega}}{B_v^4}, \Phi]
        - \nabla \cdot (\frac{P}{B_v^2}\nabla^\perp\Phi) + \nabla \cdot (\tilde{j} \boldsymbol{B}) + B_v[\frac{1}{B_v^2}, p] + \nabla\cdot({\mu_\perp\nabla^\perp\tilde{\omega}}) - \Delta^\perp \cdot(\mu_h \nabla^\perp \tilde{\omega})
    \end{equation}
    \begin{equation}
        \frac{\delta\rho}{\delta t} = -B_v[\frac{\rho}{B_v^2},\Phi] + \nabla \cdot (D_\perp \nabla_\perp \rho + D_\parallel \nabla_\parallel \rho) + S_p
    \end{equation}
    \begin{equation}
        \frac{\delta p}{\delta t} = -\frac{1}{B_v}[p,\Phi]-\gamma p B_v [\frac{1}{B_v^2}, \Phi] + \nabla \cdot [(\gamma-1)\kappa_\perp \nabla_\perp T+ (\gamma -1 )\kappa_\parallel T_\parallel + \frac{pD_\perp}{\rho}\nabla_\perp \rho + \frac{pD_\parallel}{\rho}\nabla_\parallel \rho] + (\gamma -1)(S_e + \eta_{Ohm}B_v^2\tilde{j}^2) 
    \end{equation}
    \begin{equation}
        \frac{\delta \Psi}{\delta t} = \frac{\delta^\parallel \Phi - [\Psi,\Phi]}{B_v} - \eta(\tilde{j} - \tilde{j_0}) + \nabla \cdot (\eta_n \nabla^\perp \tilde{j})
    \end{equation}
    \begin{equation}
        \tilde{j} = \Delta^*\Psi
    \end{equation}
    \begin{equation}
        \tilde{\omega} = \Delta^\perp\Phi ,
    \end{equation}
\end{subequations} \\

with operators $\delta_\parallel = B_v^{-1}\nabla\chi\cdot\nabla$, $\nabla^\perp = \nabla - B_v^{-1}\nabla\chi\delta_\parallel$, $\Delta^\perp = \nabla \cdot \nabla^\perp$, and $\Delta^* = B_v^{-2}\nabla\cdot(B_v^2\nabla^\perp$.
Note that a simplified viscosity is used in place of a tensor in the original MHD equations presented, based on the assumption that the viscosity is anisotropic \cite{nikulsin_models_2021}. This is not necessarily accurate for magnetized plasmas, but is sufficient for our present modelling purposes. This simplification also allows the viscous term to be omitted in the derivation and replaced at the end with a generic viscosity term. \\

Once the reduced equations are derived, they must be discretized in time and space for implementation in JOREK.

\subsection{JOREK numerical scheme} \label{joreknumericalscheme}
JOREK is an extended non-linear MHD code, which successfully models many large-scale plasma phenomena \cite{hoelzl_jorek_2021}. The base JOREK model is designed for tokamaks. The basis of the code is a 2D Bézier finite element scheme in the poloidal plane. The physical variables are then extended to the entire torus via a Fourier expansion. Several time discretization schemes are available, such as Crank-Nicolson, Gears, and Euler, largely to achieve different levels of numerical diffusion as required.
The starting equilibrium is calculated based on a prescribed initial pressure profile, using a solver for the Grad-Shafranov equation. The Grad-Shafranov equation is an equation for the poloidal magnetic flux, and is itself a reduction of ideal MHD equations to 2 dimensions \cite{freidberg_plasma_2007}.  The system is then evolved through time using reduced MHD. Several extensions to this exist, such as full MHD, kinetic effects, and two-fluid effects. 

\subsection{JOREK stellarator extension} \label{sec:jorek3d}
This project concerns the stellarator extension to the code.  There are a few alterations which must be made to make the code suitable to stellarators \cite{nikulsin_jorek3d_2022}. The poloidal plane varies toroidally in stellarators, so the grid in the RZ plane must be extended in the toroidal direction via Fourier expansion, similarly to the physical variables. An alternative, ansatz method is used for reducing the equations, as seen in section \ref{sec:fulltoreducedMHD}. Additionally, the Grad-Shafranov equation is not applicable to the 3-dimensional geometry, so more steps are involved to initialize the system. \\

The model is split into a first part  for calculating initial conditions and the main equations for evolving the state through time.  
Before the simulation is started, an equilibrium must be determined using an existing equilibrium solver such as GVEC \cite{nikulsin_jorek3d_2022, noauthor_vmec_nodate}. This equilibrium is then fed into the initial conditions routine, converted to the finite element basis, and the initial conditions routine solves for the initial conditions at the prescribed resolution. \\

Due to the different discretizations of GVEC and JOREK, the initial conditions do not actually constitute a perfect equilibrium. The residual forces will cause oscillations that can potentially affect the dynamics of the system. This is resolved by evolving the state initially through very small timesteps with a dissipative temporal discretization scheme, until the residual forces are sufficiently damped. This stage of simulation generally accounts for around one Alfvén time in the simulation. \\

Once the residual forces are reduced, the numerical scheme can be changed to a less dissipative and unconditionally numerically stable scheme%
, such as Crank-Nicolson, and the time step size increased to improve computational efficiency. The state is once again evolved in time, and physical perturbations can grow.  \\

\newpage

\section{Implementation and Analysis of E$\times$B flows} \label{sec:implementation}

This section will first describe the simulation set-up used throughout the project. It will then cover the steps taken to implement the source of E$\times$B rotation in the code, and briefly discuss the other methods that could have been used to implement these flows. It will also discuss the steps taken to determine whether the effects were consistent with expectations.

\subsection{The Test Case} \label{sec:testcase}

\begin{figure}[h]
    \centering
     \begin{subfigure}[b]{0.45\textwidth}
         \centering
         \includegraphics[width=\textwidth]{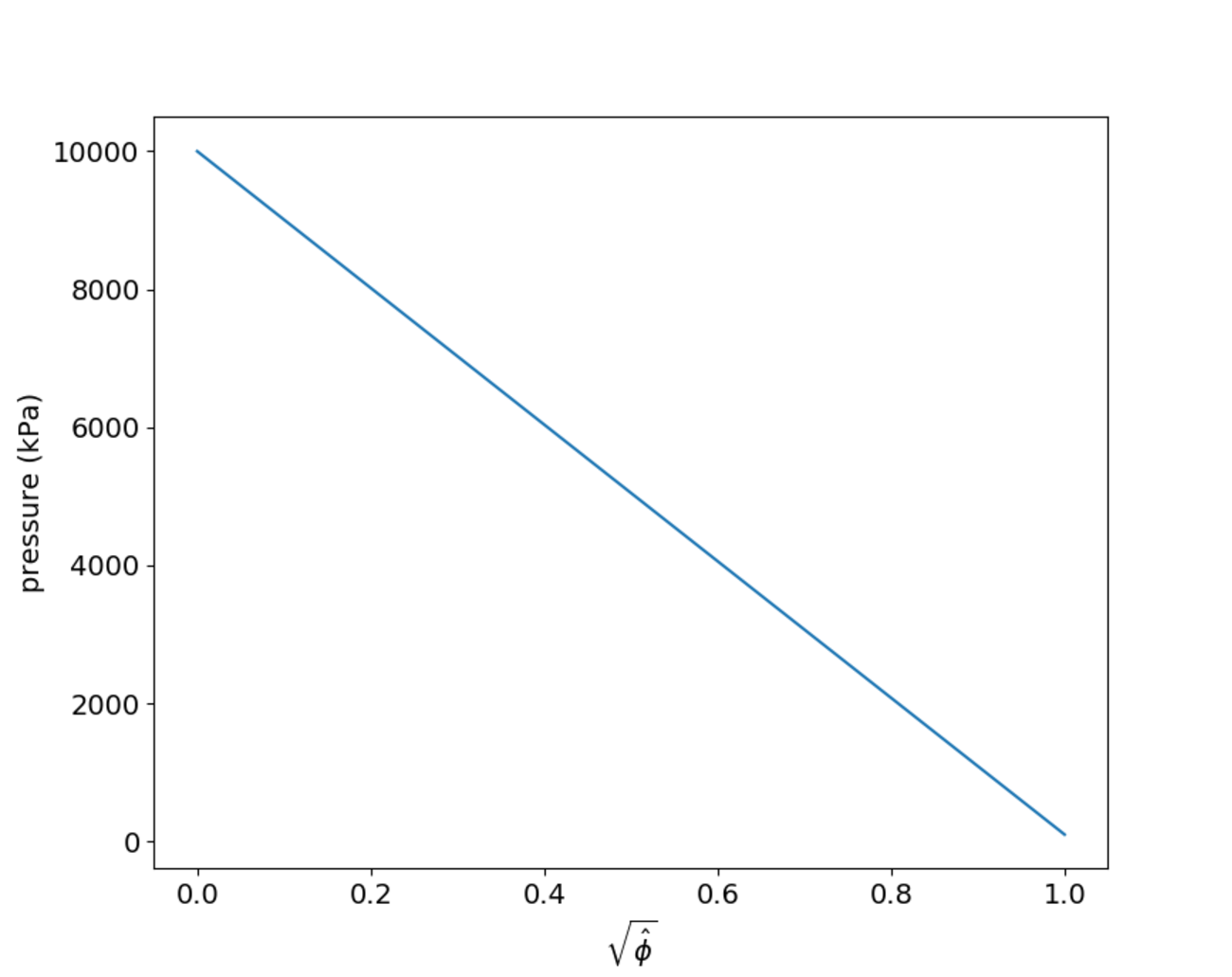}
         \caption{}         \label{fig:pressureprofile}
     \end{subfigure}
          \begin{subfigure}[b]{0.45\textwidth}
         \centering
         \includegraphics[width=\textwidth]{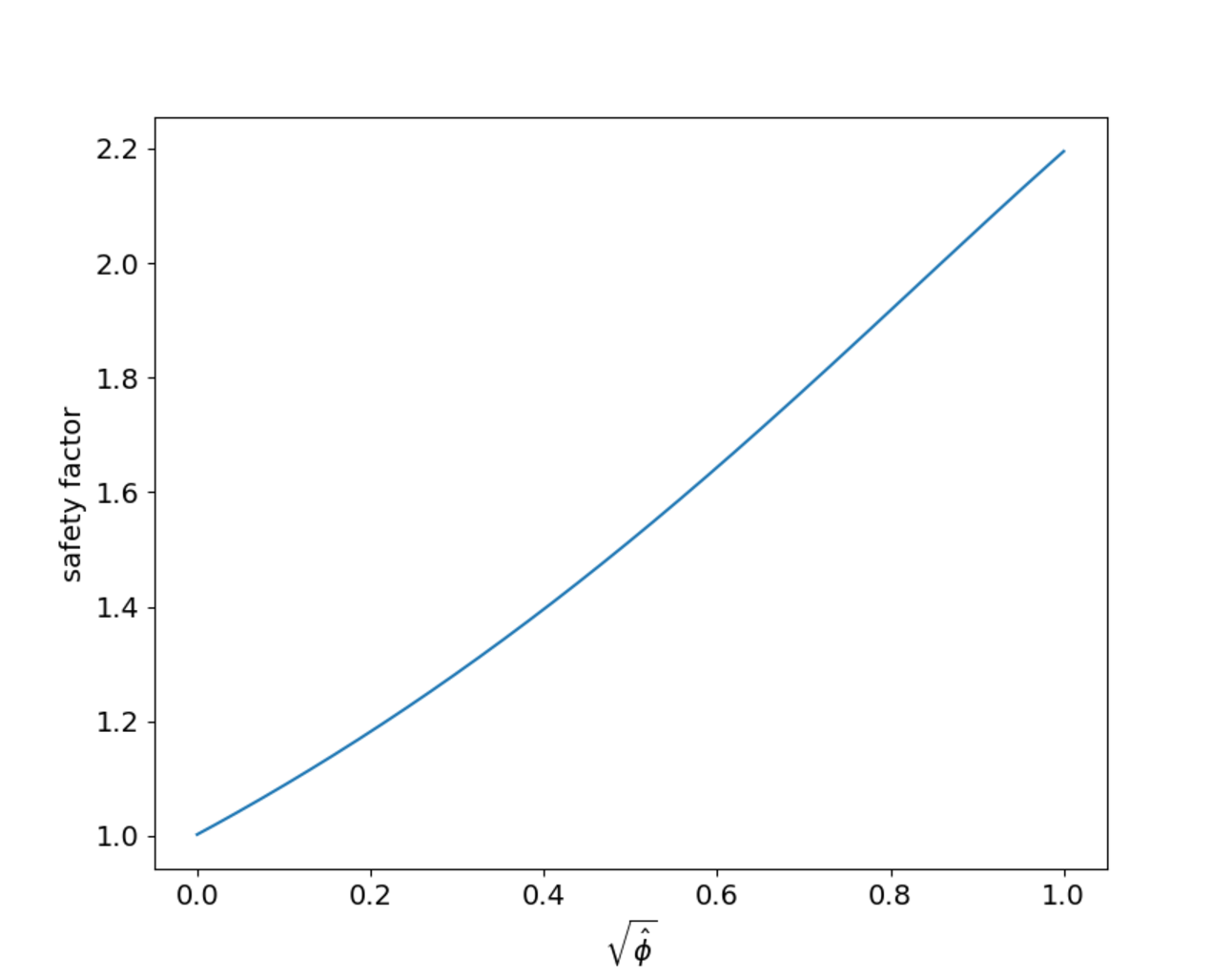}
         \caption{}         \label{fig:qprofile}
     \end{subfigure}
     \caption{a) pressure (in kPa), and b) safety factor, both as functions of the square root of the normalized toroidal flux.}
    \label{fig:enter-label}
\end{figure}

The ballooning simulation used as the basis for this project is developed in \cite{nikulsin_models_2021}. The equilibrium is based on Wendelstein 7-A stellarator geometry with a vacuum field strength varying from 2.2 to 2.5 Tesla and a pressure of 10kPa at the center of the plasma. W7-A is a classical l=2 stellarator, meaning that the plasma shape is a simple ellipse in the poloidal cross-section and formed by helical coils wrapping around the entire torus \cite{grieger_wendelstein_1985}. The case used here contains toroidal mode numbers n = 0, 5, and 10, with the intention of demonstrating effects applicable to high-n ballooning modes in general, while limiting the computational costs. We expect ballooning in these modes due to the strong pressure gradient (fig \ref{fig:pressureprofile}) combined with the high magnetic shear (fig \ref{fig:qprofile}); the magnetic shear is too large for an interchange mode to arise, but we still expect a pressure-driven instability due to the high pressures. \\

The objective is to run a simulation with low values of resistivity, viscosity, and diffusivity. However, setting these parameters too low can result in numerical instabilities unless a very high grid resolution is used, which increases the computational costs. In order to determine the maximum values of each parameter that won't effect the growth rate of the least stable (n=10) mode, parameter scans are performed. \\

\begin{figure}
     \begin{subfigure}[b]{0.45\textwidth}
         \centering
         \includegraphics[width=\textwidth]{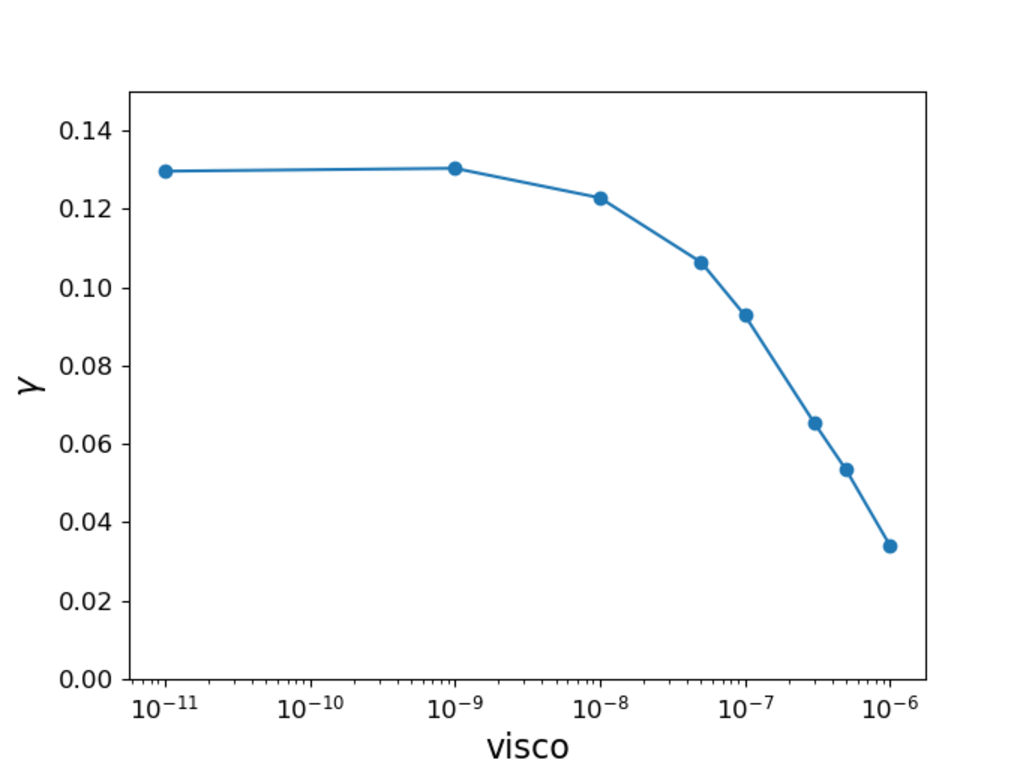}
         \caption{Average linear growth rate $\gamma$ as a function of the viscosity $\mu$.}
         \label{fig:ballooninggr_visco}
     \end{subfigure}
     \begin{subfigure}[b]{0.45\textwidth}
         \centering
         \includegraphics[width=\textwidth]{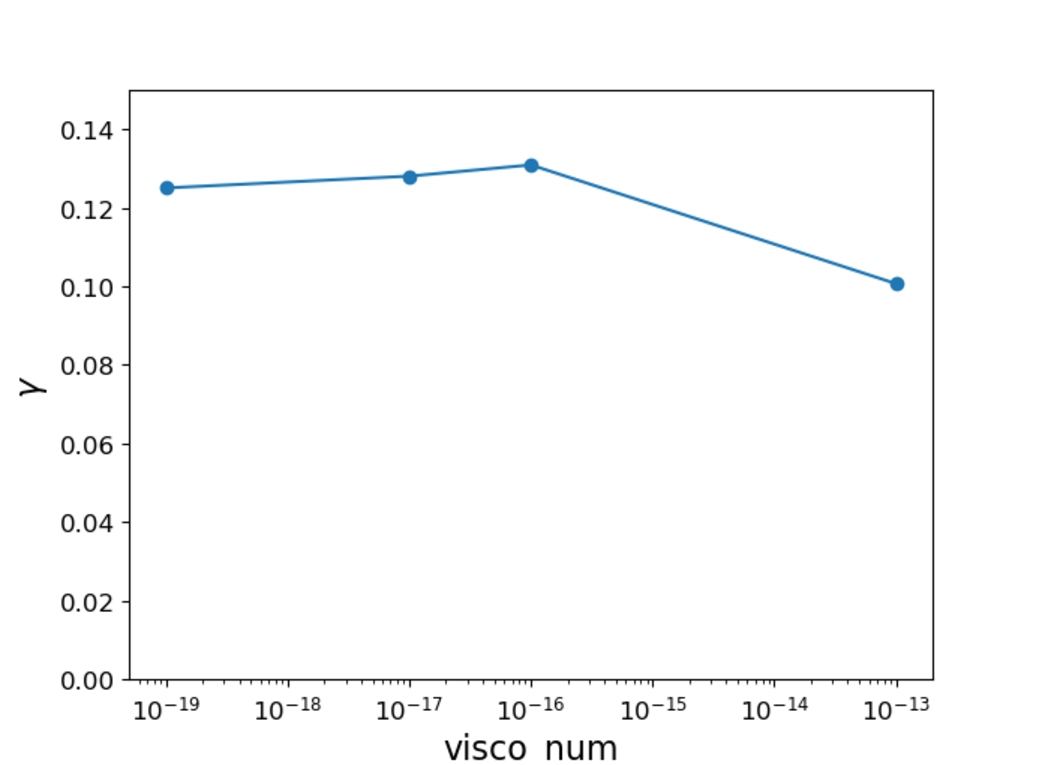}
         \caption{Average linear growth rate $\gamma$ as a function of the hyperviscosity $\mu_h$.}
         \label{fig:ballooninggr_visco_num}
     \end{subfigure}
     \begin{subfigure}[b]{0.45\textwidth}
         \centering
         \includegraphics[width=\textwidth]{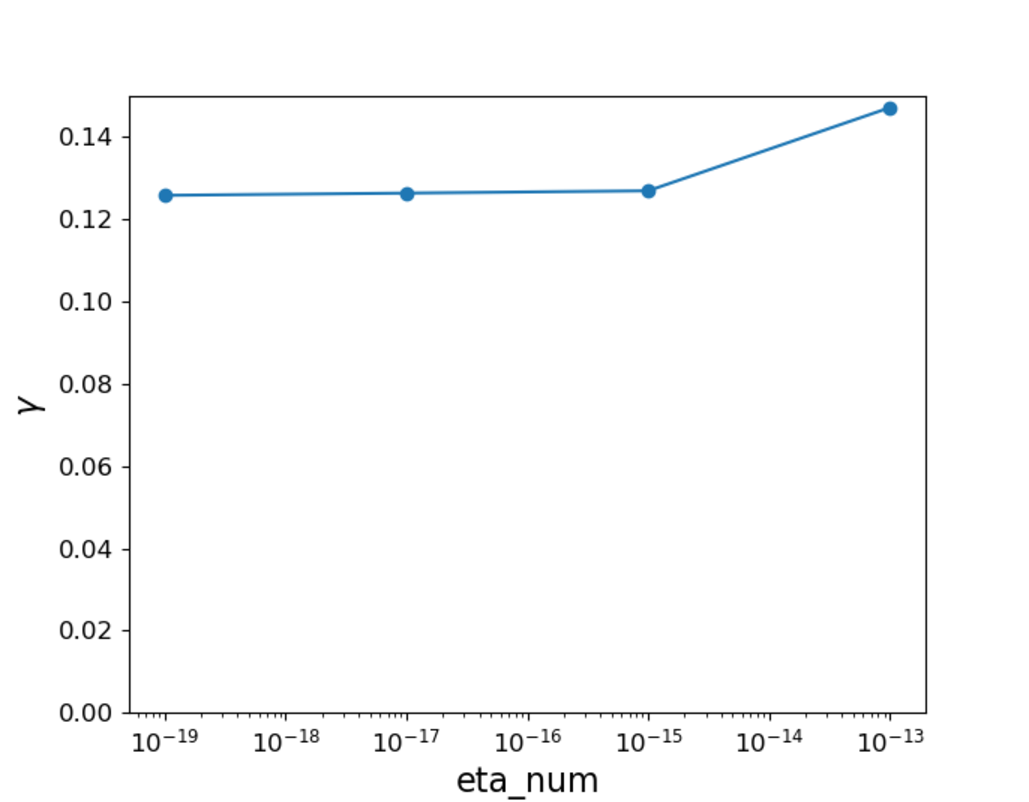}
         \caption{Average linear growth rate $\gamma$ as a function of the hyperresistivity $\eta_h$.\\}
         \label{fig:ballooninggr_eta_num}
     \end{subfigure}
     \begin{subfigure}[b]{0.45\textwidth}
         \centering
         \includegraphics[width=\textwidth]{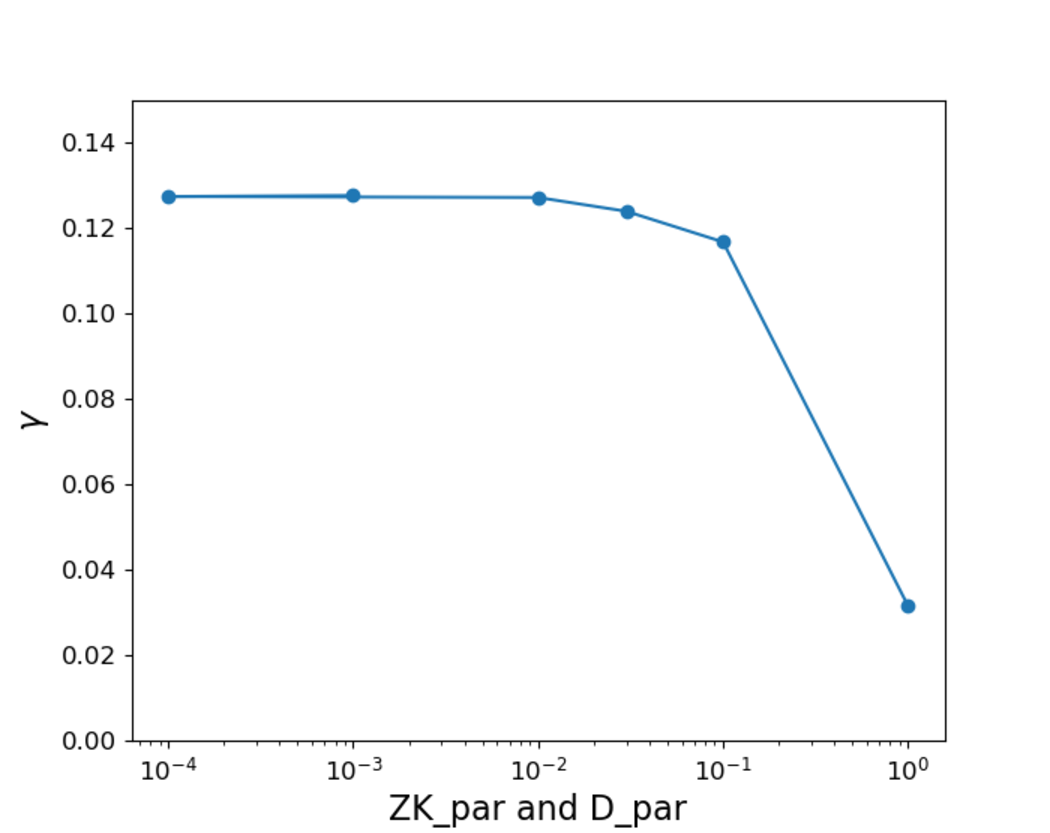}
         \caption{Average linear growth rate $\gamma$ as a function of the parallel mass diffusivity $D_\parallel$ and parallel heat diffusivity $\kappa_\parallel$. $D_\parallel = \kappa_\parallel$ for scan.}
         \label{fig:ballooninggr_par_diff}
     \end{subfigure}
     \begin{subfigure}[b]{0.45\textwidth}
         \centering
         \includegraphics[width=\textwidth]{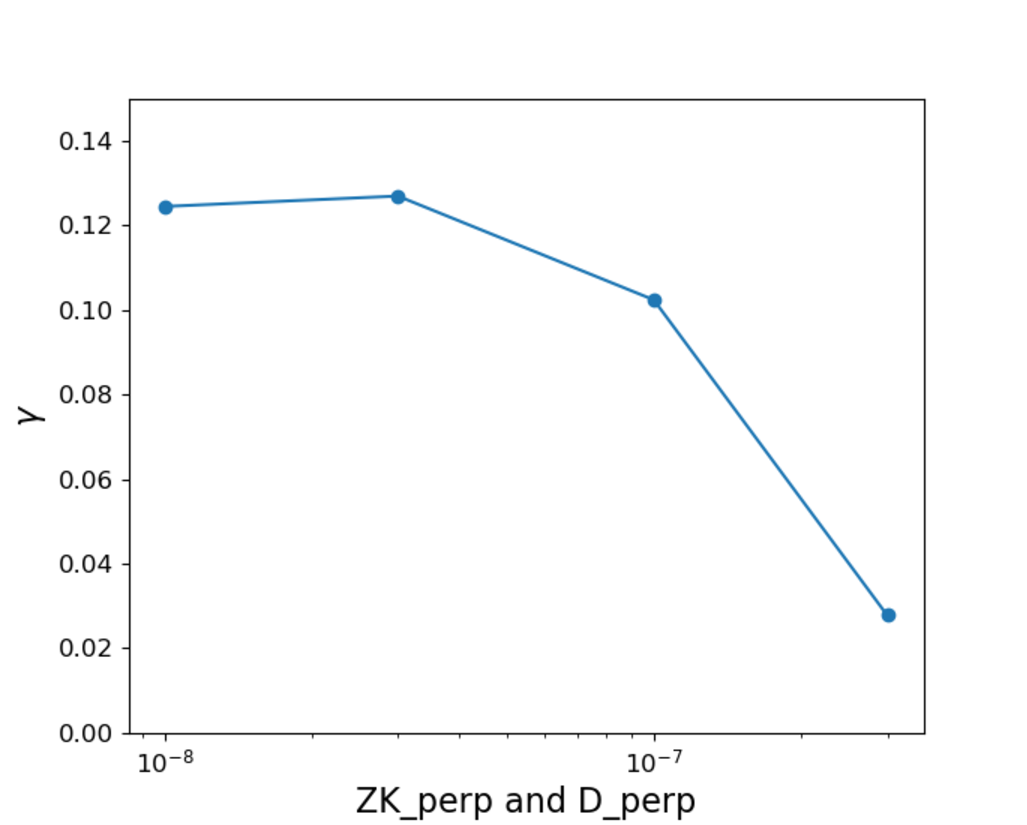}
         \caption{Average linear growth rate $\gamma$ as a function of the perpendicular mass diffusivity $D_\perp$ and perpendicular heat diffusivity $\kappa_\perp$. $D_\perp = \kappa_\perp$ for scan.}
         \label{fig:ballooninggr_perp_diff}
     \end{subfigure}
        \caption{Growth rates of various parameter scans.}
        \label{fig:ballooninggr)}
\end{figure}

The viscosity parameter scan (fig \ref{fig:ballooninggr_visco}) is consistent with the expectation that viscosity will disperse the mode and therefore limit its growth rate. JOREK also contains a hyperviscosity parameter. This is not intended to be used to affect the dynamics. Rather, it acts on higher order terms in order to prevent numerical instabilities. The hyperviscosity should be several orders of magnitude less than the viscosity to prevent those terms from becoming dominant. For higher hyperviscosities, the hyperviscosity becomes comparable to the physical viscosity and affects the dynamics of the system in a similar way. This prediction is consistent with what is observed in the hyperviscosity scan (figure \ref{fig:ballooninggr_visco_num}). \\

It is already known that ballooning mode growth rates increase with resistivity, so a scan is not performed on this parameter \cite{huysmans_external_2005}. JOREK also contains a hyperresistivity for numerical stability, which will have a similar effect as resistivity on the system if it is too large, as seen in fig \ref{fig:ballooninggr_eta_num}. In JOREK units.\\

Diffusivity of particles and heat causes decreased growth rates. This effect is more noticeable with perpendicular diffusivities because the mode is more localized perpendicular to field lines. However, it is seen that the parallel diffusivities affect the dynamics at lower values than expected. This is because the simulation uses a low number of toroidal harmonics, thus the magnetic field lines are not perfectly aligned to the flux surfaces. Therefore, the parallel diffusivity coefficient will actually contribute to perpendicular diffusivity. It is observed in figure \ref{fig:ballooninggr_par_diff} that at high enough diffusivity, the parallel diffusivity has a similar effect as perpendicular diffusivity.

Based on these scans, the values used for the actual simulations are chosen as shown in table \ref{tab:parameters}.

\begin{table}[h]
    \centering
    \begin{tabular}{|c|c|}
        \hline
         viscosity  &  $1\times10^{-9}$\\ 
         \hline
         hyperviscosity &  $1\times10^{-15}$ \\ 
         \hline
         hyperresistivity &  $1\times10^{-17}$\\ 
         \hline
         Parallel diffusivities (particle and heat) & $1\times10^{-2}$ \\ 
         \hline
         Perpendicular diffusivities (particle and heat)  & $1\times10^{-8}$ \\ 
         \hline
         
    \end{tabular}
    \caption{Viscoresistive \& diffusive parameters for the test case determined from scans. In JOREK units.}
    \label{tab:parameters}
\end{table}

\subsection{Prescribing a profile} \label{sec:prescribingprofile}
\begin{figure}
    \centering
    \begin{subfigure}[b]{0.7\textwidth} 
        \centering
        \includegraphics[height=10cm]{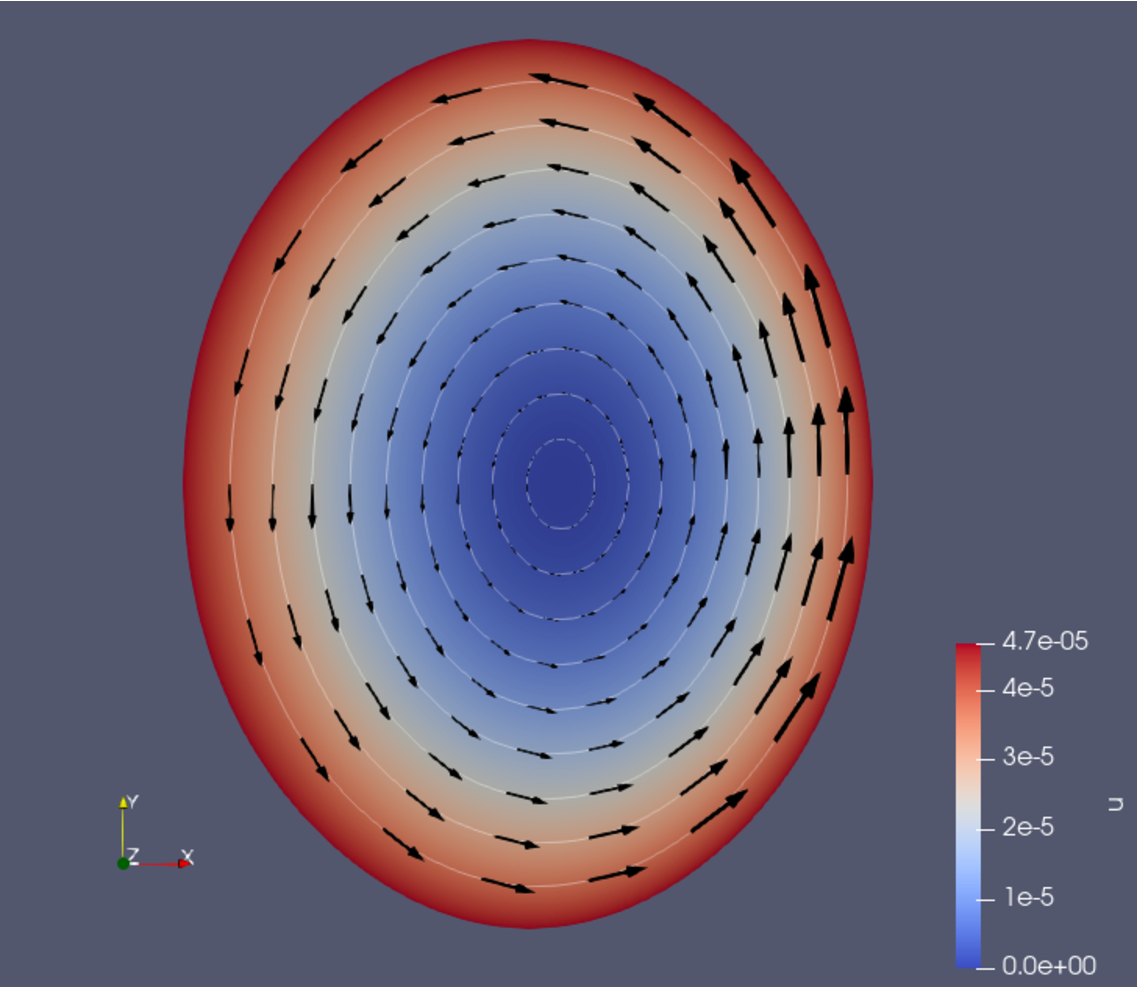}
        \caption{Poloidal cross-section showing the electric potential (colour plot), and the velocity (black arrows) shown on contours of the square root of normalized toroidal flux.}
    \label{fig:1km_poloidalcrosssection}
    \end{subfigure}
    \begin{subfigure}[b]{0.49\textwidth}
         \centering
         \includegraphics[width=\textwidth]{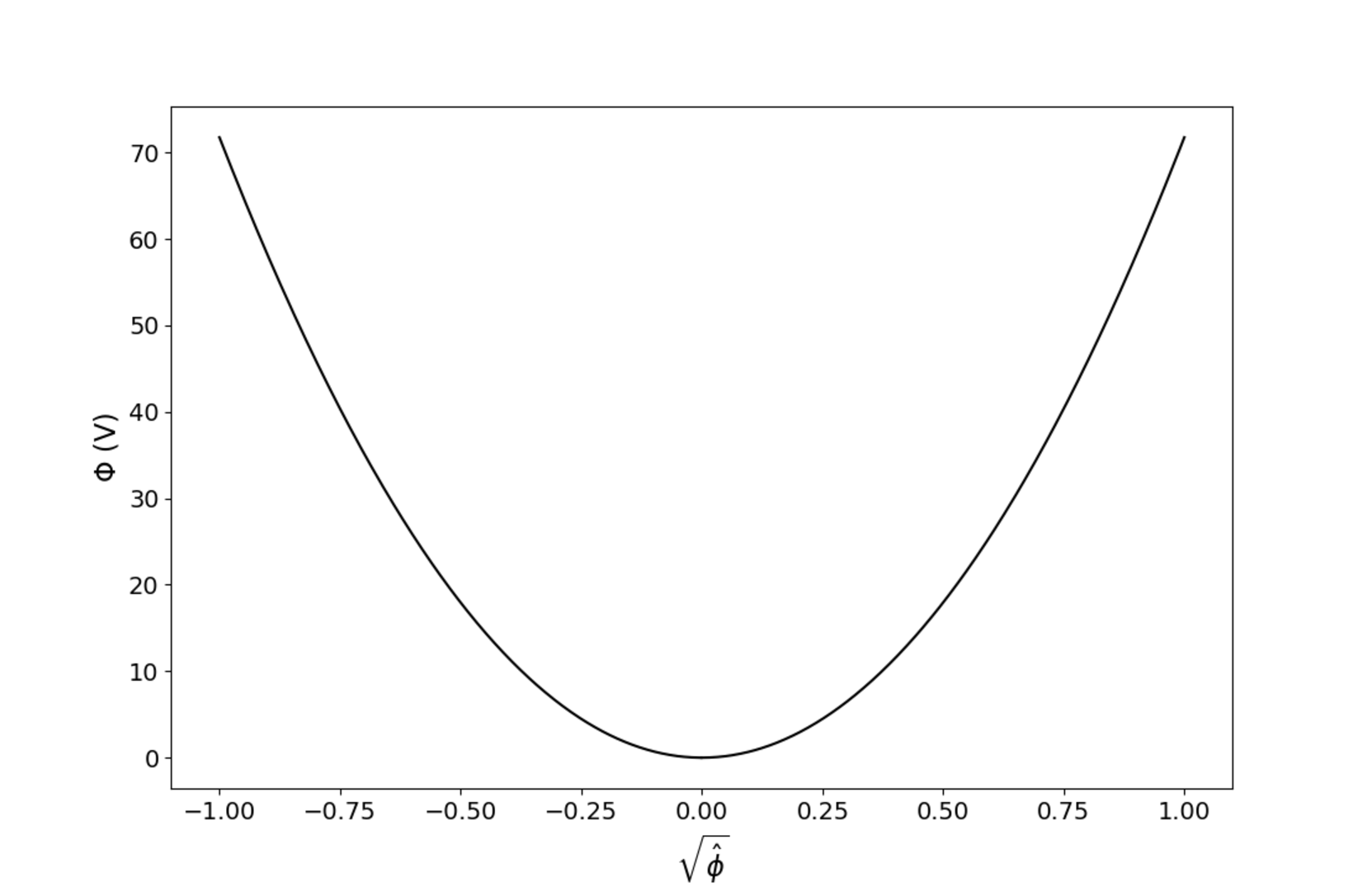}
         \caption{The electric potential as a function of the square root of the normalized toroidal flux.\\}
         \label{fig:1kmPhi}
     \end{subfigure}
     \begin{subfigure}[b]{0.49\textwidth}
         \centering
         \includegraphics[width=\textwidth]{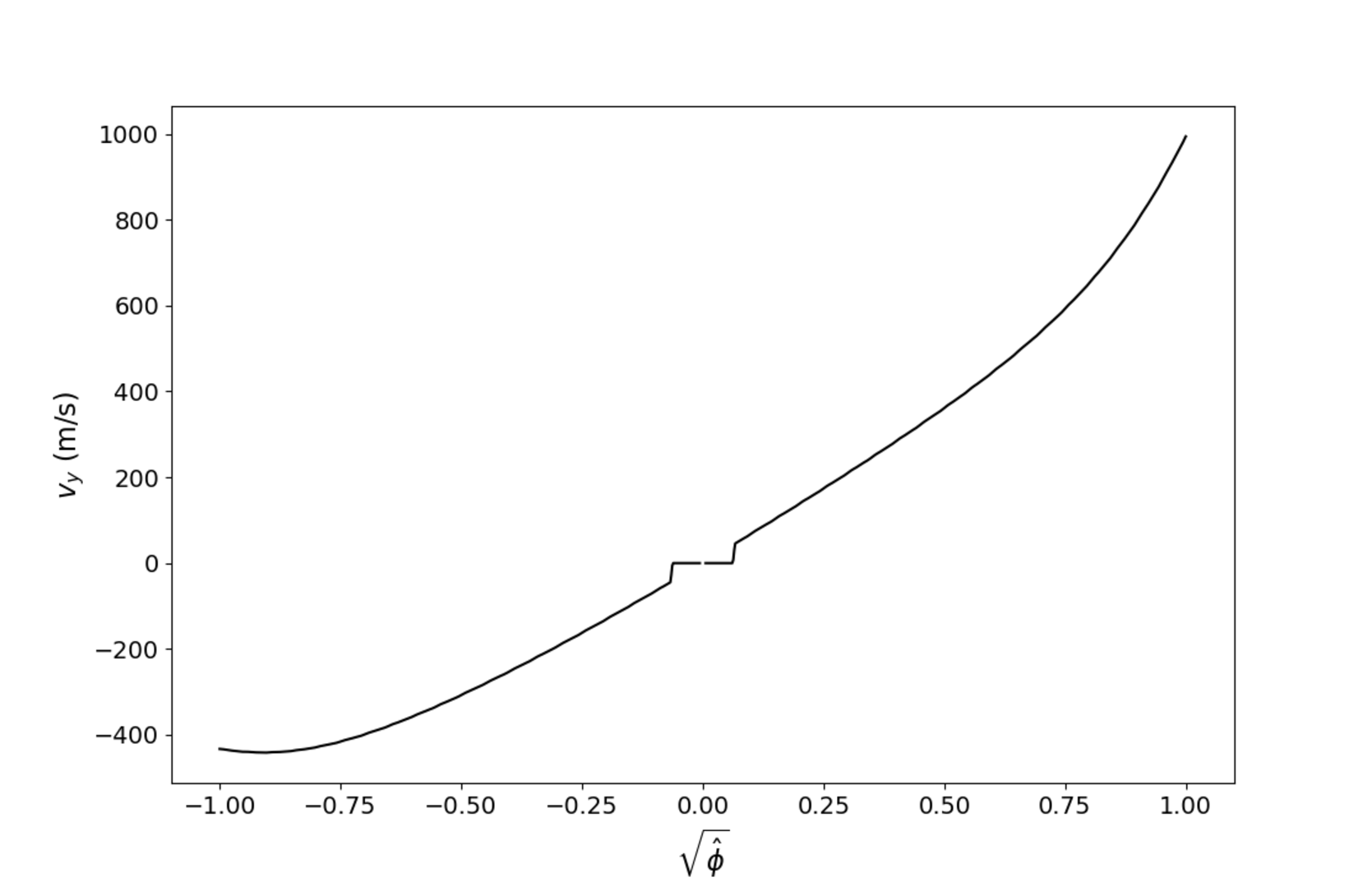}
         \caption{$v_y$ (coordinates shown in figure \ref{fig:1km_poloidalcrosssection} ) as a function of the square root of the normalized toroidal flux, at the midplane.}
         \label{fig:1kmvel}
     \end{subfigure}
        \caption{Initial conditions of the case with $v_{max}$=1 km/s.}
        \label{fig:1km}
\end{figure}

There are several ways that poloidal flows could come to be implemented in the code. As discussed in section \ref{subsec:poloidalflows}, E$\times$B drift is one of most significant contributors to poloidal flows in reactors. Due to the physical relevance of E$\times$B drift and since the MHD model already includes the presence of E$\times$B flows, as indicated by the velocity ansatz (eq. \ref{eq:velocityansatz}), it was determined that a background electric potential would be implemented as the source of flows. Since the vacuum field is primarily toroidal, a radial electric field gives a poloidal E$\times$B drift, as seen in figure \ref{fig:1kmvel}.\\

In order to see the effects of the poloidal flow, this background electric potential should be constant in time. The most straightforward way to achieve this is to simply prescribe the potential profile at the beginning of the simulation, making no changes to the time evolution equations. Taking into consideration the low values of the viscous and diffusive parameters in table \ref{tab:parameters} and the fact that the test case without poloidal flows reaches the end of the linear stage in under 200 normalized time units, we expected that the potential profile would not change significantly over the course of the simulation. Stabilization of the mode results in a longer time to reach the nonlinear phase, but this is a rather short time frame regardless. \\

The potential profile was prescribed at the beginning of the simulation, at the same time that the GVEC equilibrium is imported to JOREK. GVEC finds the minimum of the energy function to determine the equilibium state, making several assumptions including nested flux surfaces and lack of plasma flow. These assumptions guarantee that the $\nabla \Psi$ component of the momentum equation is elliptic (where $\Psi$ is the poloidal flux), thus providing a procedure to solve it \cite{hameiri_equilibrium_1983}.  Introducing flows into the system introduces the possibility that the momentum equation will be hyperbolic, rendering the equilibrium solution not strictly valid. It can be shown that this only occurs at Mach numbers exceeding unity \cite{hameiri_equilibrium_1983}. Thus flows which are below the ion sound speed can be implemented without problems. \\

\subsection{Choice of potential profile}

Any choice of potential profile could be implemented this way. For the purposes of testing, a parabolic function of the square root of the normalized toroidal flux was chosen. 

This is a reasonable first test to maintain numerical stability; the profile is not particularly steep anywhere, and avoids discontinuities in the flow profile. \\ 

The parabolic potential profile and resulting velocity field are shown in figure \ref{fig:1km}, for a case with a maximum poloidal velocity of 1km/s. When different flow velocities are simulated in this project, the same profile is used, but with an additional scaling factor. Note that the velocity in figure \ref{fig:1kmvel} is shown as being 0 near the magnetic axis; this is not representative of the actual velocity, but is rather a choice made in the diagnostic of the velocity to avoid diagnostic issues when the Jacobian approaches 0. The velocity is determined by equation \ref{eq:ecrossbdrift}, but the scaling of each part of the equation must be considered. The square root of the normalized toroidal flux $\sqrt{\hat{\phi}}$ scales approximately with the minor radius $r$. The mapping of the square root of the normalized toroidal flux to the minor radius is distorted by the Grad-Shafranov shift, a common phenomenon in toroidal devices in which the plasma pressure causes flux surfaces to be shifted outwards; $\frac{d\sqrt{\hat{\phi}}}{dr}$ is therefore greater on the outboard side of the reactor. The rotation of the poloidal cross section with the toroidal coordinate is also a consideration. The poloidal cross-section shown in figure \ref{fig:1km_poloidalcrosssection} is elongated in the vertical direction, thus the contribution of the $\frac{d\sqrt{\hat{\phi}}}{dr}$ term is lesser than it is the horizontal direction. The term will be different at different toroidal locations.
Additionally, the scaling of the vacuum field with 1/R results in a linear scaling of velocity with the major radius. \cite{hoelzl_jorek_2021}. \\

Following the implementation of the profile and initial testing, it was discovered that the prescribed electric potential does change significantly throughout the simulation, rendering the flow stabilization hard to interpret. This was resolved by not updating the n=0 component of the potential profile in the time evolution. This is acceptable as long as the n=0 mode does not contribute much to the instability, which is true for the case at hand but may not be valid for all cases.

\subsection{Alternative solution} \label{subsec:alternaltivesolution}
An alternative solution would be to introduce a source term for the potential into the time-evolution equation of the potential. This would be introduced as a momentum source to the MHD momentum equation, and then have the projections applied to it, resulting in a source term with the form 

\begin{equation}
    \nu \cdot [f(\boldsymbol{\Phi}_0)-f(\boldsymbol{\Phi})]
\end{equation}
where $\boldsymbol{\Phi}_0$ is the imposed potential, $\boldsymbol{\Phi}$ is the potential at the given time step, and $\nu$ is a viscous coefficient. This would prevent the potential profile from varying too much from what is prescribed, while allowing the n=0 component to contribute to the instability if that is what the physics dictates. This solution is more robust, however the function f is rather complicated in the reduced MHD formulation. This solution was outside the scope of the project due to the limited timescale.

\subsection{Measures of flow stabilization}
As discussed in section \ref{subsec:poloidalflows}, there are multiple ways that plasma flows can cause stabilization. Some consideration is needed to parse out the different effects. 

\subsubsection{Shearing rate} \label{subsec:shearingrate}

Sheared velocity flows are expected to stabilize ballooning modes by shearing apart the mode structures. The amount of shear can be measured via the shearing rate, defined as

\begin{equation} \label{shear}
    \omega_{s} = |\frac{\Delta  v_{\theta}}{\Delta r_\theta}| 
\end{equation}
with $\Delta v_\theta$ being the difference in poloidal velocity across the radial width of the mode and $\Delta r_\theta$ being the poloidal length scale of the mode. \\

To get the average shear over the mode, we need to measure the width of the mode for the case without flows, in the poloidal and radial directions. This is performed on the low-field side of the plasma, where the mode is strongest, and in the latter stage of linear growth where the mode structure can be be discerned from the perturbations to background fields. \\

The poloidal width of the mode is obtained by plotting the temperature across a flux surface within the mode, in the linear phase of a case with no flows. The arclength from the minimum to maximum temperature $2\Delta \theta$ is measured, then halved and converted to meters to obtain a characteristic length scale $\Delta r_\theta$ for the mode. \\

The shearing rate varies over the radial direction, but we are interested in the average shear in the region where the mode appears. This is calculated by analyzing the radial velocity profiles, at the midplane on the outboard side of the plasma. This is where the mode is most noticeable, and also simplifies calculations since cylindrical coordinates can be used, with the vertical velocity being equal to the poloidal velocity at this point. The location of the ballooning mode can be observed from a fluctuation of the radial velocity. Before the mode appears, there is no radial flow. When the mode does appear, it causes radial flow since the field lines are frozen-into the flux surfaces in an ideal mode. The fact that the radial flow is only inwards at a constant poloidal location indicates that the mode is ideal. In a resistive mode, the flow would be inwards and outwards at different radial points at a constant poloidal location, since the field lines can move towards each other and reconnect.\\

Taking the width of the fluctuation of radial velocity away from noise level then gives the radial width of the mode. The poloidal velocity is then measured on either side of this determined width. The quotient of these values, $\frac{\Delta v_\theta}{\Delta r_\theta}$ is then the average shearing rate across the mode. \\

\begin{figure}
    \centering
    \begin{subfigure}[b]{0.45\textwidth} 
        \centering
        \includegraphics[height=5cm]{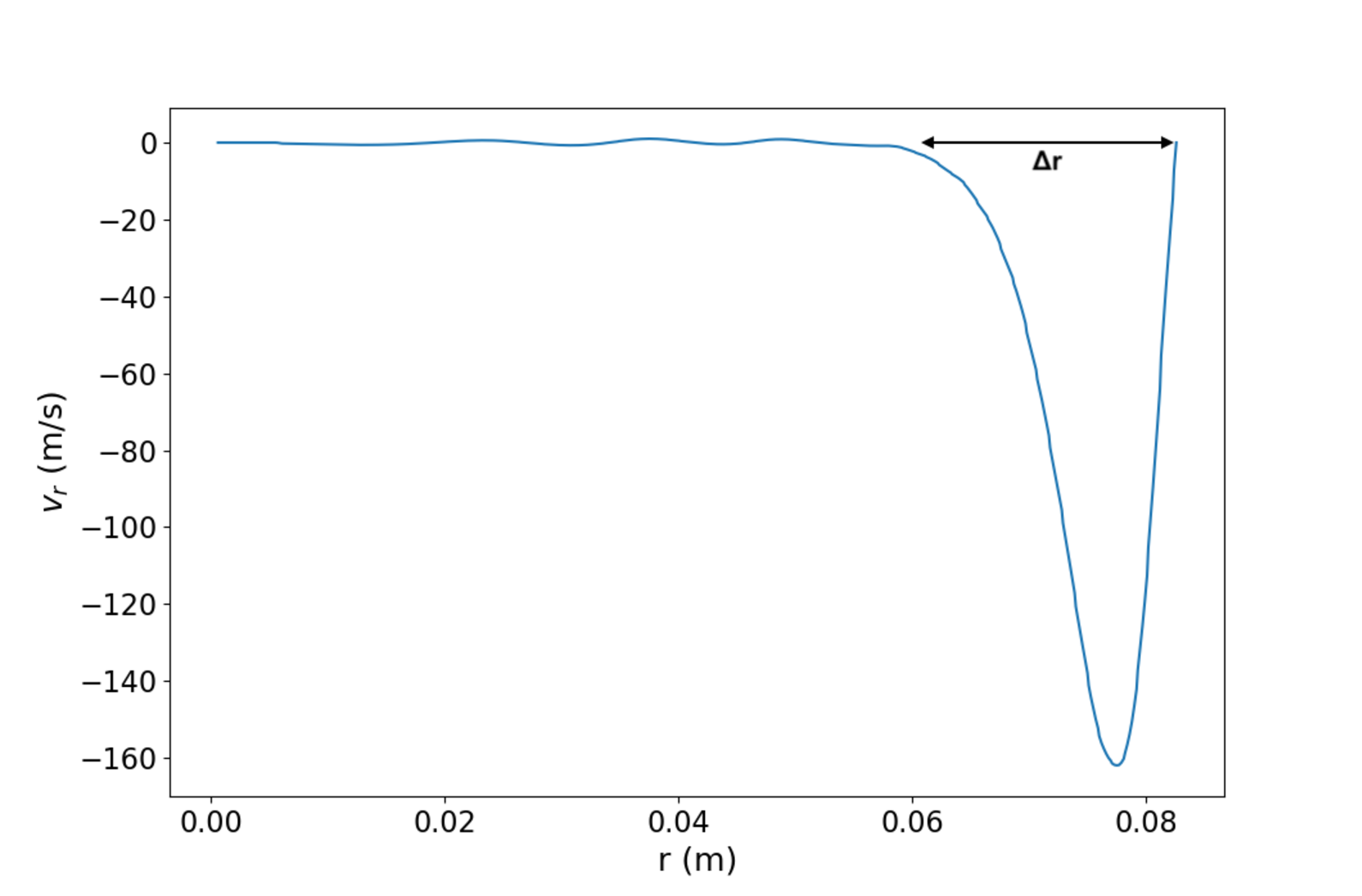}
        \caption{Radial velocity $v_r$ as a function of the minor radius r. The radial width of the mode $\Delta r$ is indicated by the black arrow. \\}
    \label{fig:deltarad}
    \end{subfigure}
    \hspace{0.5cm}
    \begin{subfigure}[b]{0.45\textwidth} 
        \centering
        \includegraphics[height=5cm]{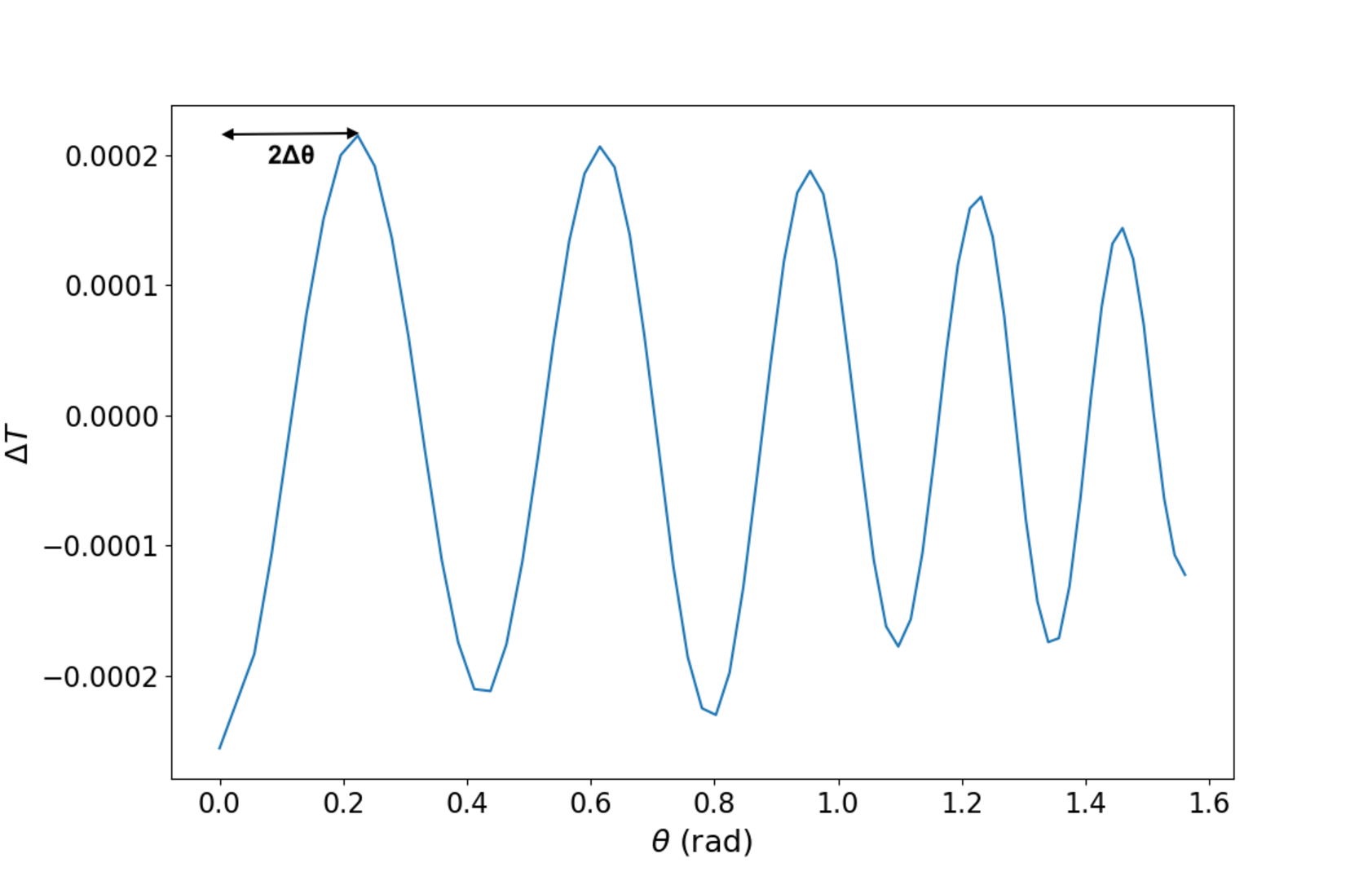}
        \caption{n=10 temperature perturbation $\Delta T$ as a function of the poloidal coordinate $\theta$. The arclength representing the poloidal width of the mode $2\Delta \theta$ is indicated by the black arrow.}
    \label{fig:2deltatheta}
    \end{subfigure}
    \caption{Measurements of the length scales for calculating shear, taken from the case without flows (see section \protect \ref{sec:basecase}) at $\hat{t}$=155.}
\end{figure}

The shearing rate needs to be compared to the average linear growth rate, which is taken as

\begin{equation}\label{avglingr}
    \gamma =  \frac{ln(\frac{E(t_2)}{E(t_1)})}{2(t_2 - t_1)}, 
\end{equation}

with E being the energy for time points $t_1$ and $t_2$ at the beginning and end of the linear phase, respectively. 
Stabilization of the n=10 mode is expected around the the point where the shear rate across the mode is equal to the n=10 growth rate of the case with no flows.

\subsubsection{``Rigid" rotation} \label{subsec:rigidrotation}
Non-sheared flow can also affect the stability by moving the plasma between areas of favourable and unfavourable curvature. If the mode was rotated with a rigid plasma flow, it would be expected to see stabilization of the mode, or potentially a growth rate that oscillates at the frequency of rotation. 
As seen in figure \ref{fig:1km}, the chosen velocity varies in both the radial direction and the poloidal direction. Since the rotation of the plasma is not rigid, there are some challenges with discerning mechanisms of stabilization that aren't due to the shear. \\ 

One could examine the effect of a non-sheared poloidal flow by producing a velocity field with no to little shear at the relevant flux surfaces, but this would need consideration to avoid sharp, difficult-to-resolve gradients, especially near the center; this is outside the scope of the project. \\

Additionally, the velocity is not constant on flux surfaces. In this project, rotation is characterized by calculating average poloidal velocity along individual flux surfaces in a given poloidal plane. Here, the surfaces $\sqrt{\hat{\phi}}$ = 0.7 and $\sqrt{\hat{\phi}}$ = 0.99 are chosen, since these are located at the inner and outer edges of the mode. The period of oscillation is the quotient of perimeter of the flux surface and the average velocity on the flux surface. This period can then be compared with oscillations in the energies of the modes in time (see section \ref{sec:oscillationofgrowthrate}).

\newpage

\section{Results} \label{sec:Results}
In this section, the appearance of ballooning modes in the base (no poloidal flow) case will be discussed and the effects of adding flows will be shown. Analysis of the cause of stabilization, as explained in section \ref{sec:implementation}, will be shown and discussed.

\subsection{Base Case} \label{sec:basecase}
\begin{figure}[h]
    \centering
    \includegraphics[width=0.85\textwidth]{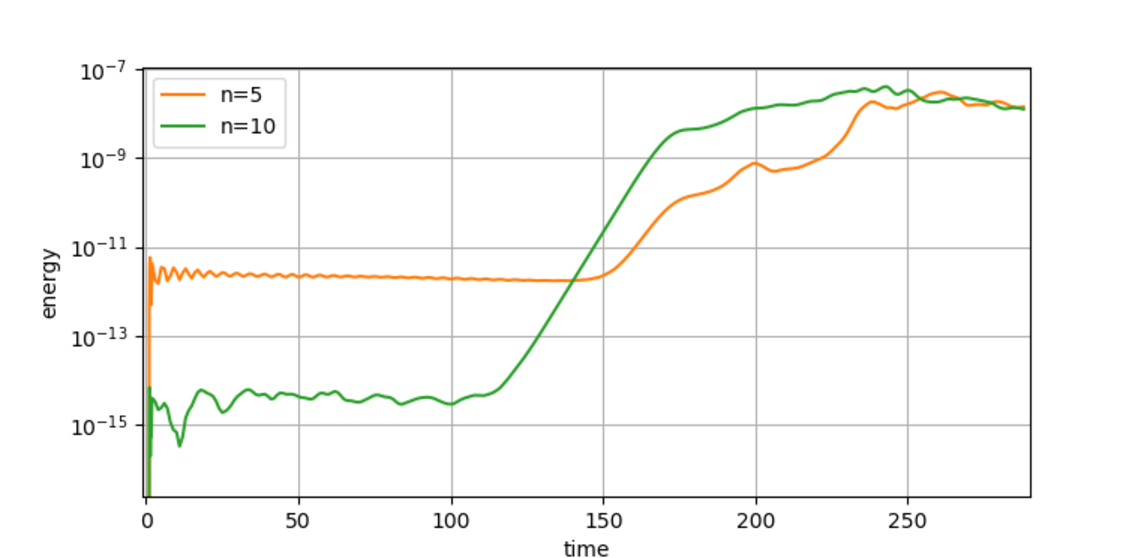}
    \caption{n=10 and n=5 kinetic energies over time, in the case without flows.}
    \label{fig:base_evt}
\end{figure}

\begin{figure}
    \centering
    \begin{subfigure}[b]{0.4\textwidth} 
        \centering
        \includegraphics[width=\textwidth]{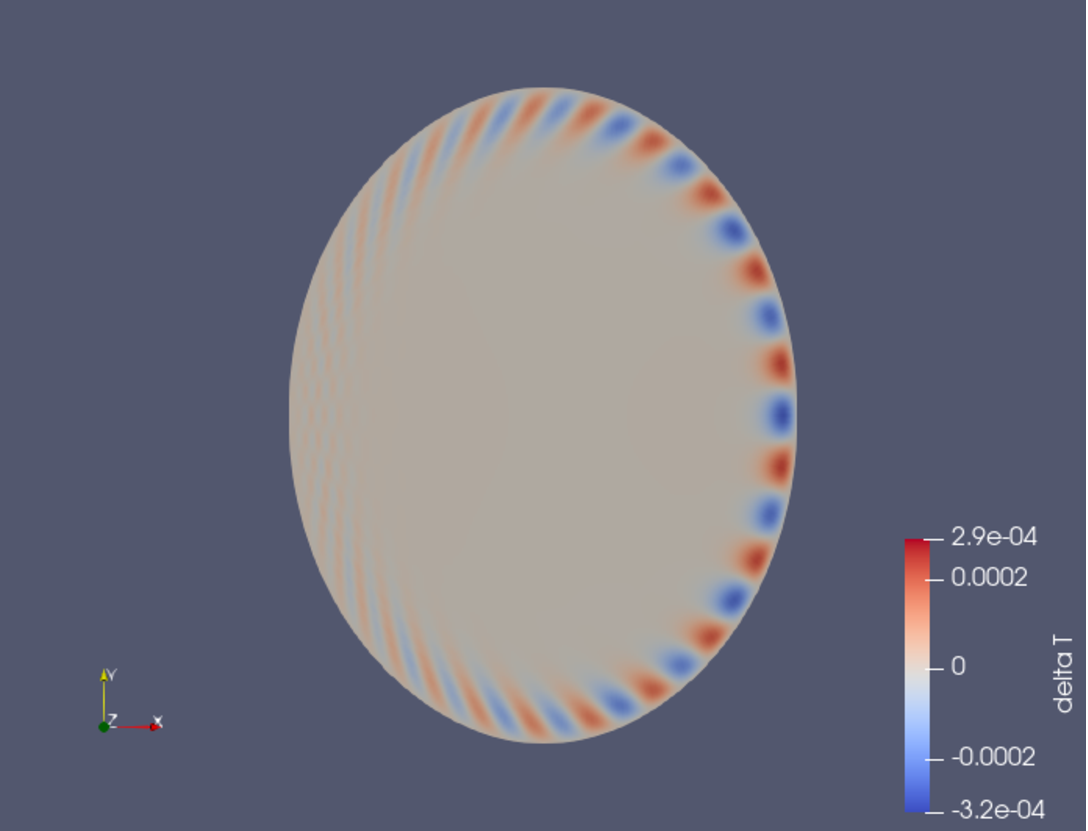}
        \caption{$\varphi = 0$}
        \label{fig:varphi0}
    \end{subfigure}
        \begin{subfigure}[b]{0.4\textwidth} 
        \centering
        \includegraphics[width=\textwidth]{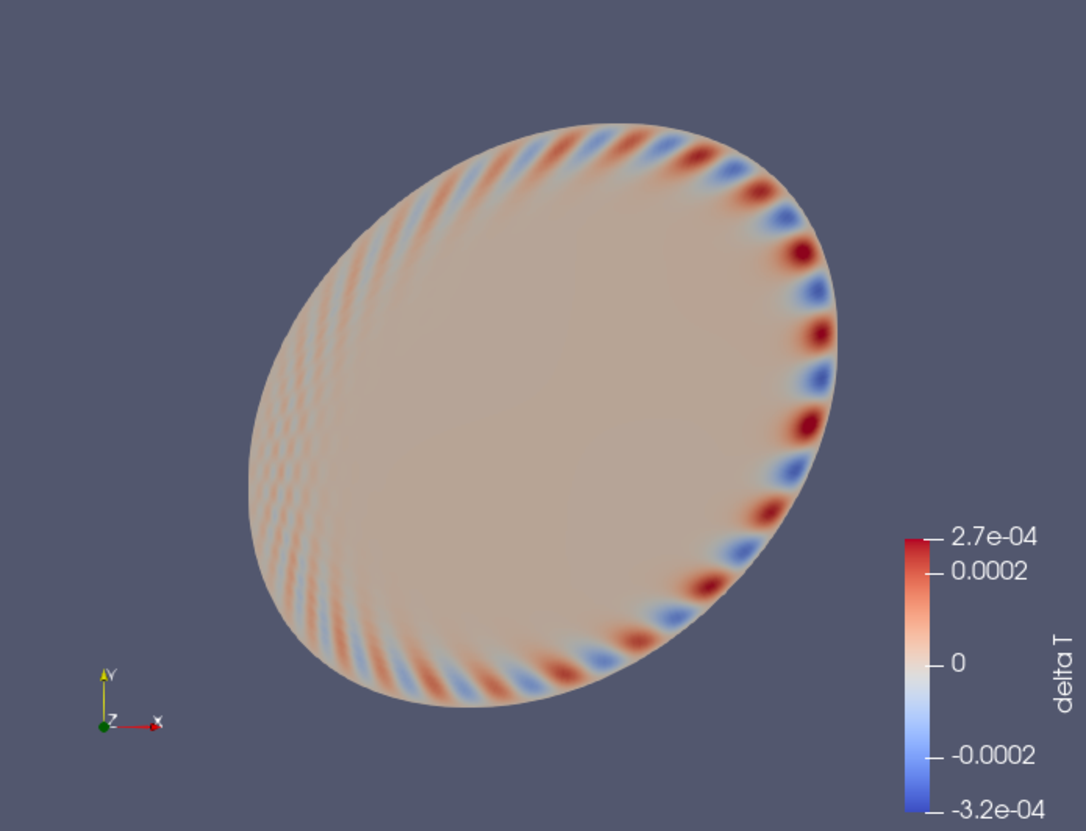}
        \caption{$\varphi = 0.31$}
    \end{subfigure}
        \begin{subfigure}[b]{0.4\textwidth} 
        \centering
        \includegraphics[width=\textwidth]{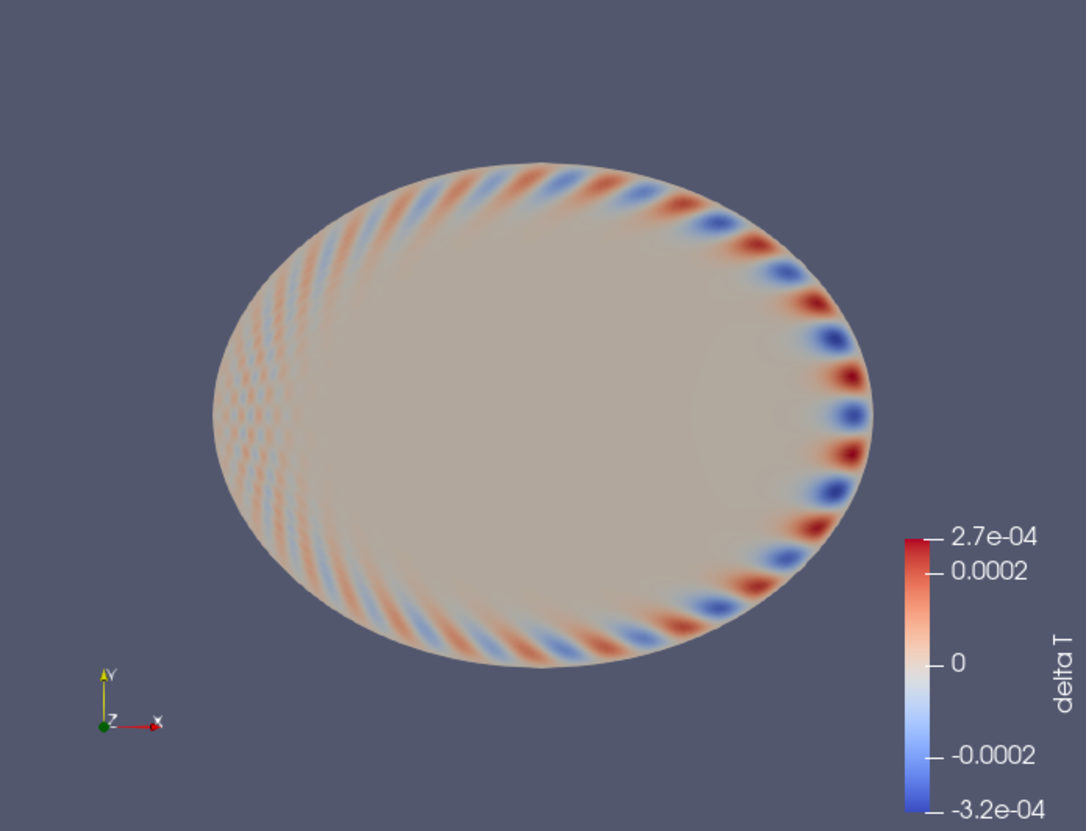}
        \caption{$\varphi = 0.63$}
    \end{subfigure}
        \begin{subfigure}[b]{0.4\textwidth} 
        \centering
        \includegraphics[width=\textwidth]{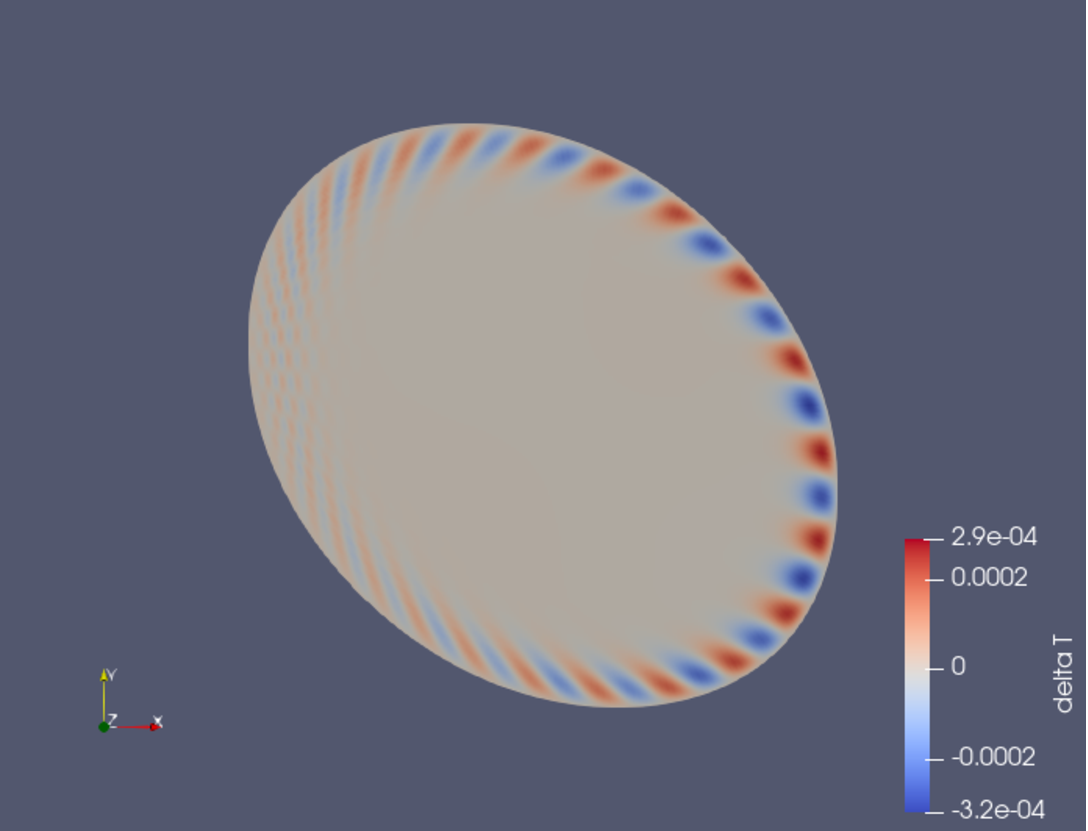}
        \caption{$\varphi = 0.94$}
    \end{subfigure}
    \caption{Poloidal cross sections at various toroidal locations $\varphi$. The colour maps show the deviation of temperature from the initial equilibrium temperature, for the case without flows at the end of the linear phase ($\hat{t}$=155).}
    \label{fig:deltatemp}
\end{figure}

The behaviour of the ballooning mode in the case without flows in shown in fig \ref{fig:base_evt}. The n=10 energy starts growing first, at around $\hat{t}$=115. The n=5 mode begins growing around $\hat{t}$=150. The linear phase ends around $\hat{t}$=175, indicated by the energies levelling off and then changing somewhat erratically past this point (non-linear phase). To fully resolve non-linear dynamics, a full Fourier spectrum would be required, which is beyond the scope of this project. The structure of the mode can be seen as deviation of the temperature from before the growth of the mode, as seen in figure \ref{fig:deltatemp}. Note that the mode appears more strongly at the low-field side of the reactor, as is characteristic of a ballooning mode. Comparatively, an interchange mode would appear evenly around the edges of the cross-section (see section \ref{sec:interchangemodes}). \\ 

\begin{figure}
    \centering
    \begin{subfigure}[b]{0.49\textwidth} 
        \centering
        \includegraphics[width=\textwidth]{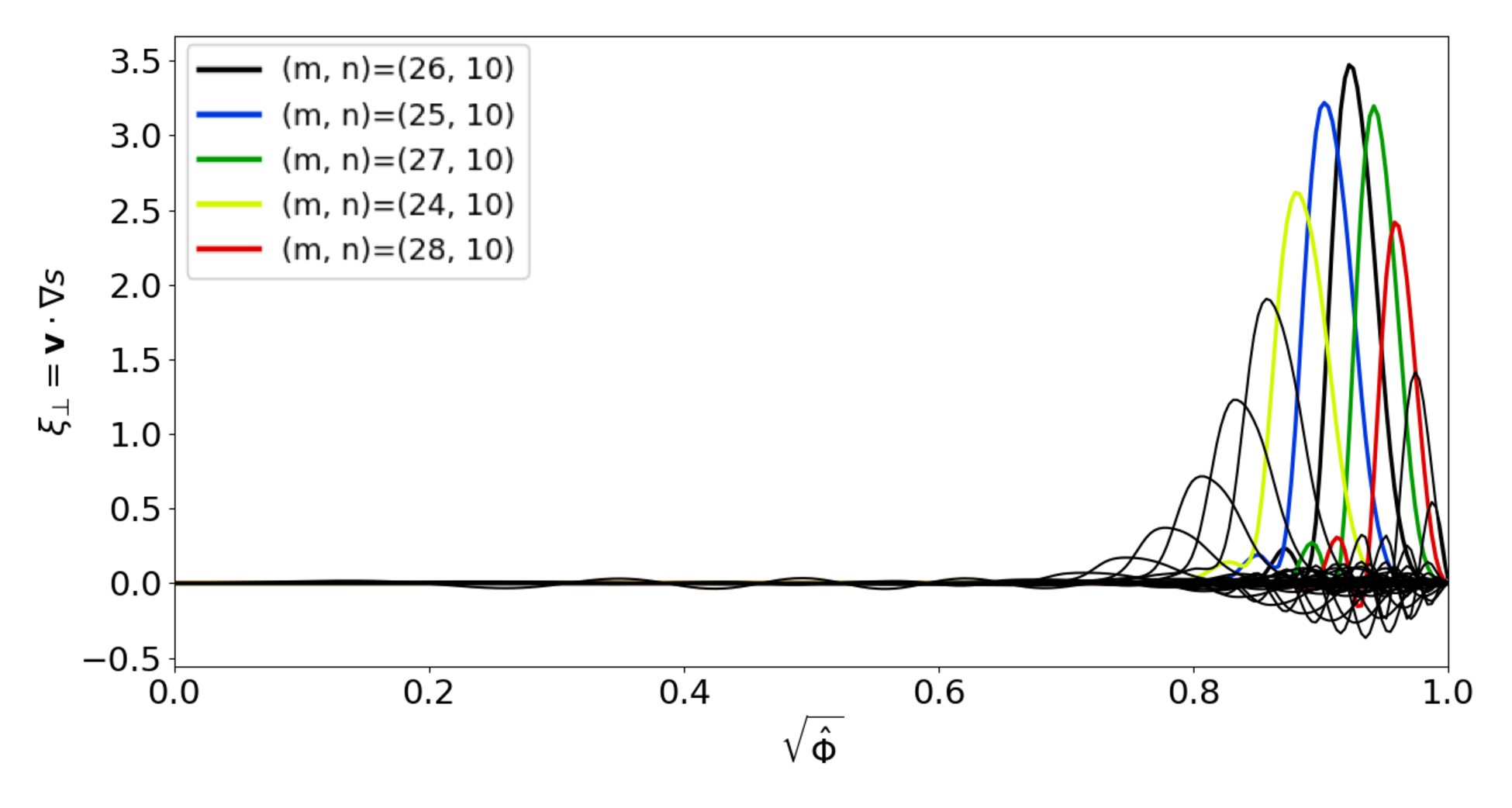}
        \caption{The real components of the dominant radial eigenfunctions over the square root of the normalized toroidal flux.}
        \label{fig:polmodesreal}
    \end{subfigure}
        \begin{subfigure}[b]{0.49\textwidth} 
        \centering
        \includegraphics[width=\textwidth]{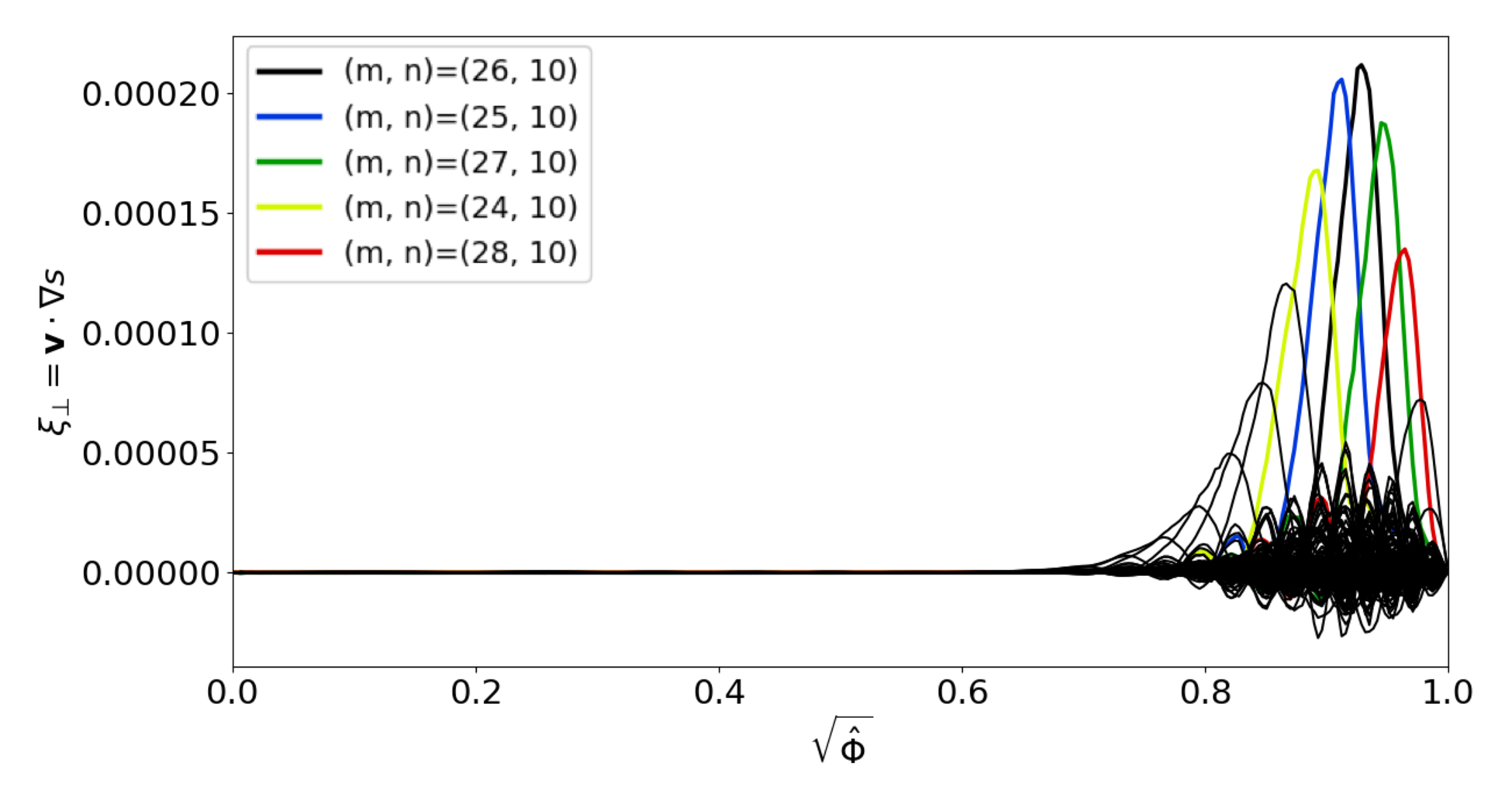}
        \caption{The imaginary components of the dominant radial eigenfunctions over the square root of the normalized toroidal flux.}
        \label{fig:polmodesaimag}
    \end{subfigure}
    \caption{Real and imaginary componenents of the dominant radial eigenfunctions near the end of the linear phase ($\hat{t}$ = 155) in the case without flows. Note that the imaginary components are negligible in comparison to the real components. This indicates that the modes all have the same phase.}
    \label{fig:polmodes}
\end{figure}

In figure \ref{fig:polmodes}, the dominant poloidal modes of the instability are plotted over the radius, at the end of the linear phase of growth. It can be seen that they overlap at the plasma edge, which causes the mode growth. The real and imaginary parts of the modes are plotted separately in figure \ref{fig:polmodes} and \ref{fig:polmodesaimag}. Recall that the phase of a mode is the arctangent of the quotient of the imaginary and real parts. It can be seen that the imaginary parts are near zero, so the modes are in phase with one another, and thus construct the mode structures observed in fig \ref{fig:deltatemp}.\\

\subsection{Stabilization}
\begin{figure}[h]
    \centering
    \includegraphics[width=\textwidth]{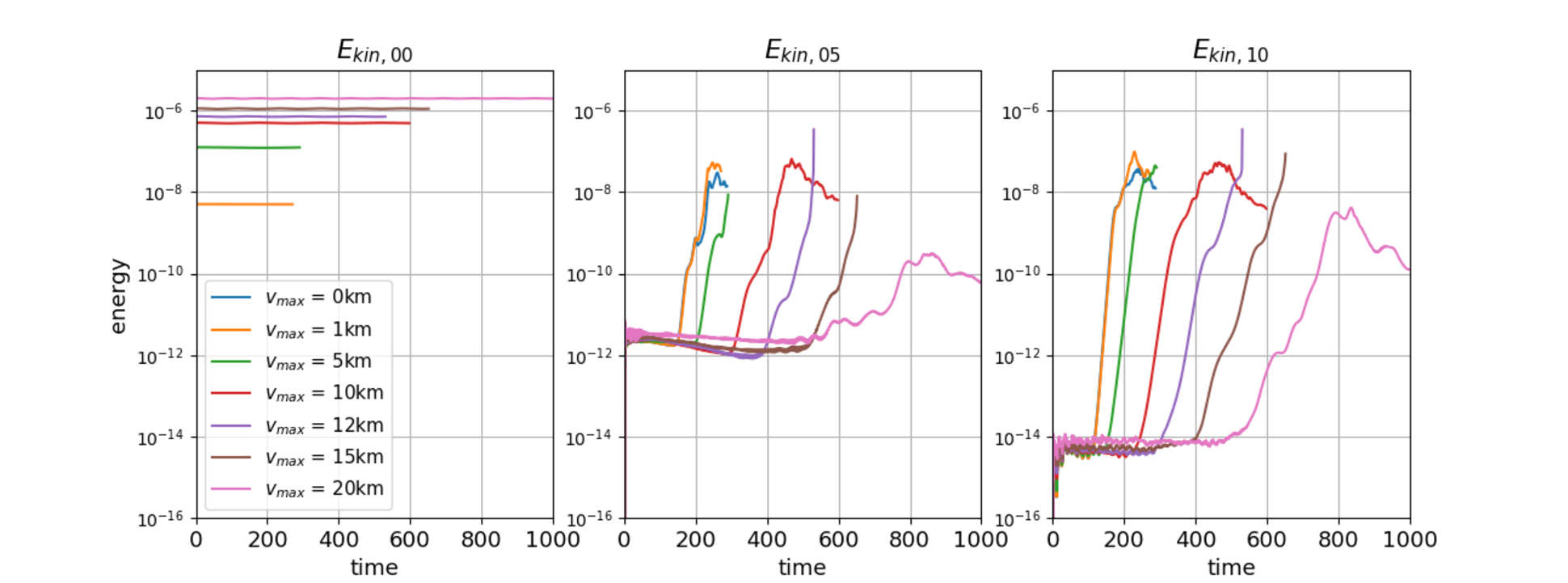}
    \caption{The kinetic energies of the n=0,5,10 modes (left to right) over time. In JOREK units.}
    \label{fig:evstime}
\end{figure}

As discussed in section \ref{subsec:poloidalflows}, we expect to see reduced growth rate with increasing poloidal flow. Figure \ref{fig:evstime} shows the kinetic energy over time for several cases of different maximum flow speed, as calculated at the low-field side of the poloidal cross-section in the toroidal plane $\varphi=0$ shown in figure \ref{fig:varphi0}. The n=0 energy increases with increasing flow speed, as expected, and remains constant. The decreasing slope of the n=5 and n=10 kinetic energies for tests with higher flows indicates that the modes are being stabilized, as predicted. \\

\begin{figure}[h]
    \centering
    \includegraphics[width=0.8\textwidth]{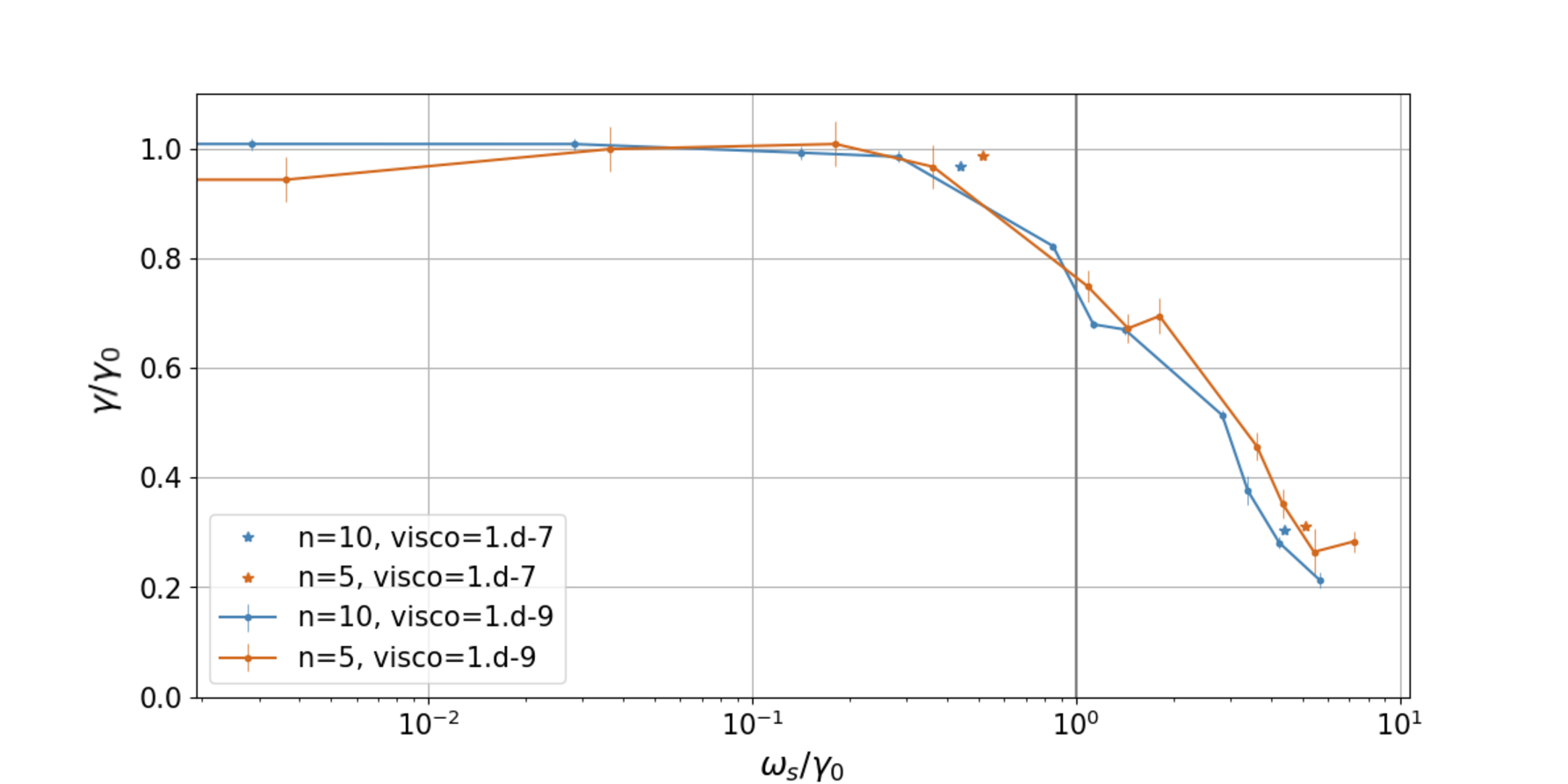}
    \caption{Average linear growth rate $\gamma$ over shearing rate $\omega_s$, both normalized to the average linear growth rate without flows $\gamma_0$. As all four curves fall together well in this normalized representation, and the stabilization occurs around $\frac{\omega_s}{\gamma_0} \approx 1$, the hypothesized stabilization by shear seems to be observed.}
    \label{fig:grvsshear}
\end{figure}

\subsection{Effects of Shear}
To observe the effect described in \ref{subsec:shearingrate}, the average linear growth rate of the mode should be compared against the shearing rate for cases with increasing flows. It can be seen that the mode is stabilized when the shearing rate is comparable to the growth rate without flows, which is consistent with expectations. This is true for both the n=5 and n=10 modes. To further support this result, additional tests were run with viscosity increased from $1\times10^{-9}$ to $1\times10^{-7}$ in JOREK units, decreasing the growth rate. Figure \ref{fig:grvsshear} shows that the stabilization occurs when the shearing rate is comparable to the growth rate without flows, regardless of factors affecting the growth rate without flows. The hypothesized mechanism of stabilization via shear flow decorrelation is thus strongly supported. This result is consistent with a similar mechanism of stabilization via sheared toroidal flows that has been studied both analytically and numerically in tokamaks \cite{cooper_ballooning_1988, waelbroeck_ballooning_1991}. \\

\subsection{Oscillation of Growth Rate} \label{sec:oscillationofgrowthrate}
\begin{figure}
    \centering
    \includegraphics[width=0.7\textwidth]{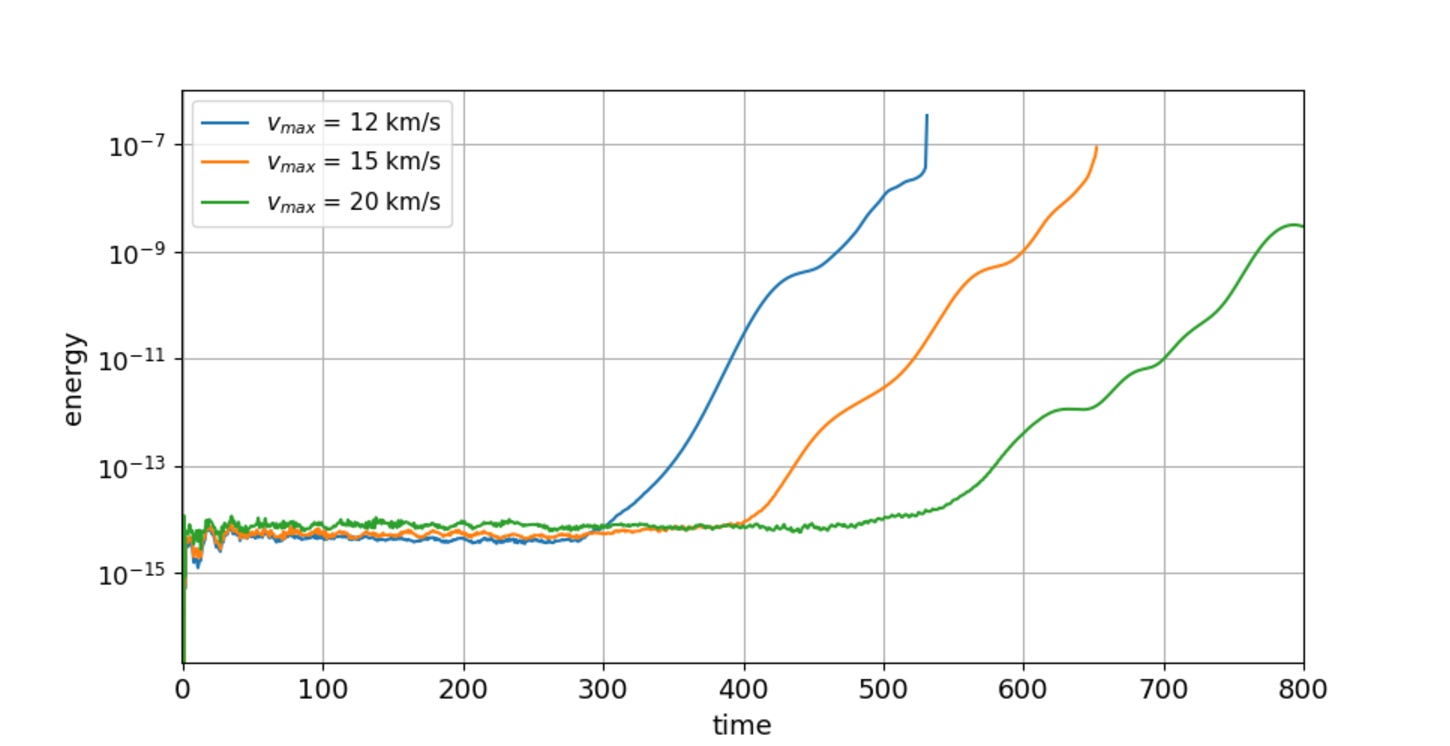}
    \caption{n=10 componenent of kinetic energy over time, for cases with changing growth rate in the linear phase.}
    \label{fig:oscillations}
\end{figure}

Looking at figures \ref{fig:evstime} and \ref{fig:oscillations}, there is an oscillation of the linear growth rate in cases with higher flow velocities. The period between maximum slope was measured for each case, and compared with the average period of rotation as described in section \ref{subsec:rigidrotation}. Table \ref{tab:periodofrotation} shows the approximate period of oscillation of the linear growth rates for the cases where such an oscillation is visible in the energy plots (fig \ref{fig:evstime}). We can then compare these values to the period of rotation, the calculation for which is described in section \ref{subsec:rigidrotation}.  It can be seen that the period of oscillation of the growth rate is not completely consistent with the expected period if this were due to convection of the mode from low- to high-field side of the plasma. The oscillation of the growth rates may be due to the rigid rotation from the low-field side to the high-field side of the plasma, but further investigation is required. \\

\begin{table}[h]
    \centering
    \begin{tabular}{|c|c|c|}
        \hline
         \bf{$v_{max} (m/s)$} & \bf{\small{average period of rotation ($\sqrt{\hat{\phi}}$ = 0.7/0.99)}} & \bf{\small{period of n=10 growth rate oscillation}} \\
         \hline
         12 & 143/132 & 125 \\
         \hline
         15 & 114/106 & 92 \\
         \hline
         20 & 86/80 & 86 \\
         \hline
    \end{tabular}
    \caption{Average period of rotation at edges of mode and period of n=10 growth rate oscillation, for cases with visible oscillation of growth rate.}
    \label{tab:periodofrotation}
\end{table}

The changing linear growth rate could have a source other than the rigid rotation. It was then hypothesized that the increasing flow speed could cause the relative phases of the poloidal modes to change over time, resulting in the oscillation of the growth. The phase difference would then be expected to be about constant in the cases without oscillations in the growth rate, and oscillate in the cases with oscillations. To test this, the dominant modes were identified and their phases were plotted over over time. Recall that in the case without flows, the modes have the same phase (fig. \ref{fig:polmodes}).\\

\begin{figure}
    \centering
    \begin{subfigure}[b]{0.48\textwidth}
        \centering
        \includegraphics[width=\textwidth]{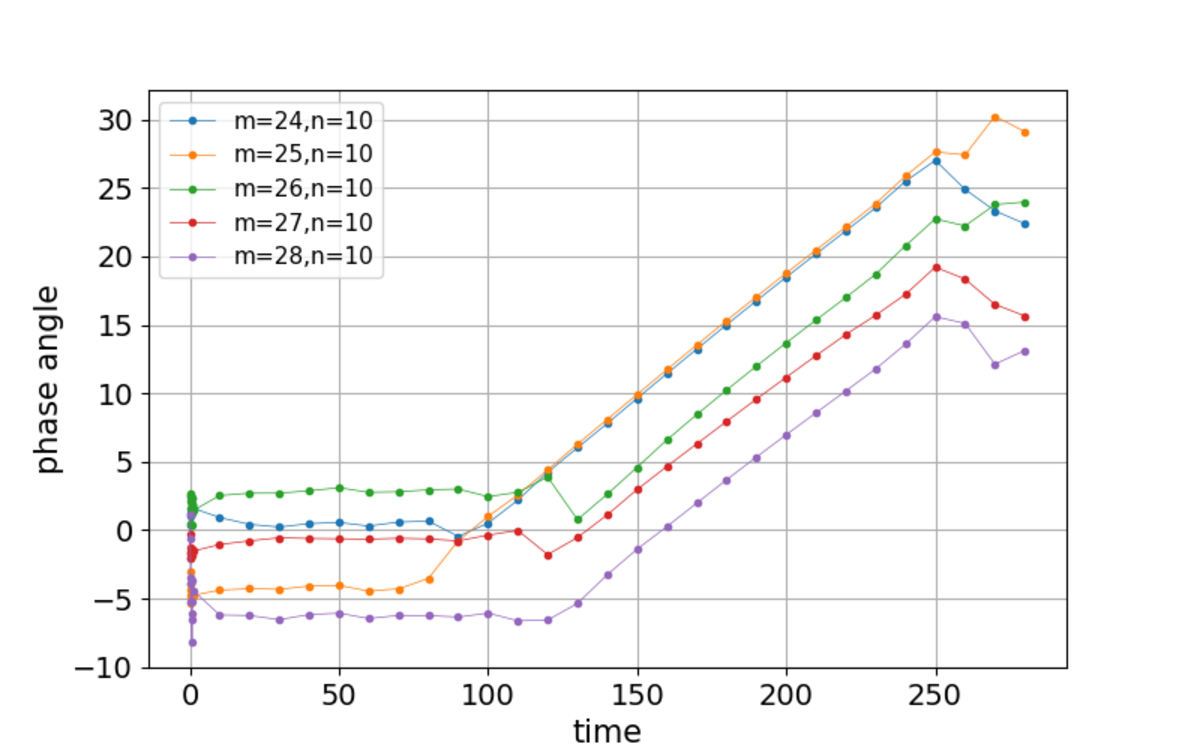}
        \caption{$v_{max}$ = 5 km/s}
    \end{subfigure}
    \begin{subfigure}[b]{0.48\textwidth} 
        \centering
        \includegraphics[width=\textwidth]{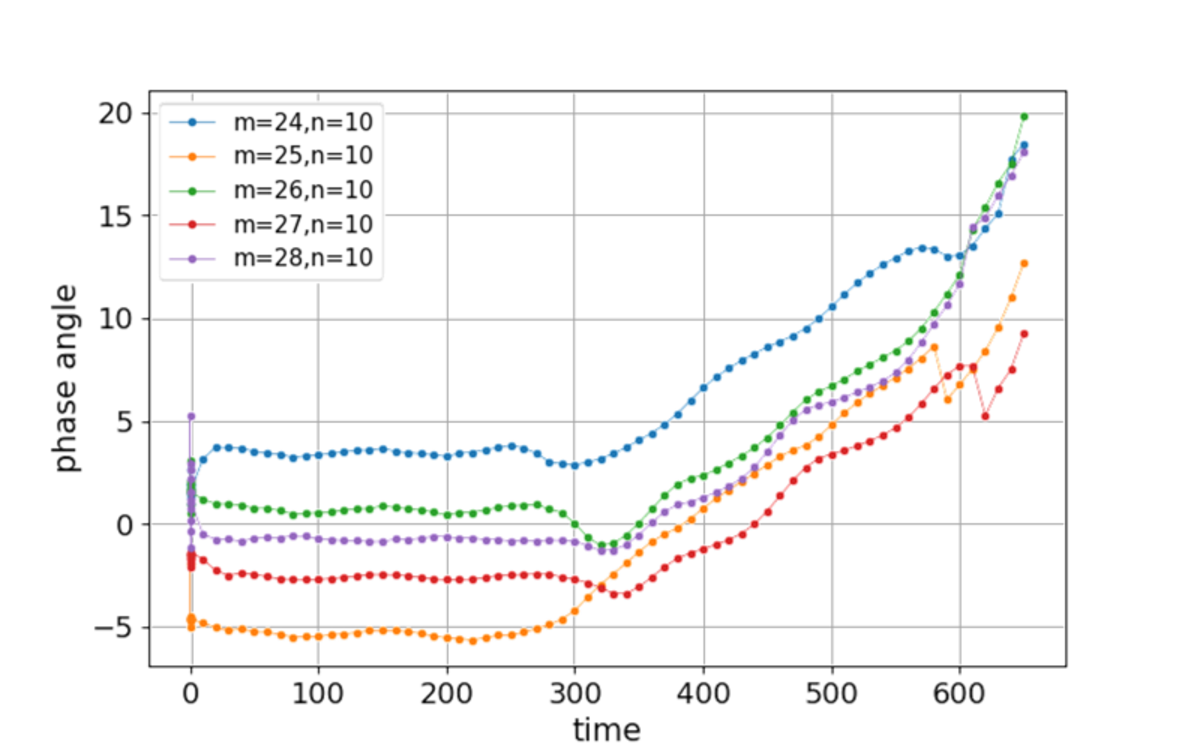}
        \caption{$v_{max}$ = 15 km/s}
    \end{subfigure}
    \caption{Phase angles of the dominant modes over time for the a) 5 km/s case, where oscillations of the growth rate is not present and b) the 15 km/s case, where oscillations of the growth rate are observed. While the phases of all modes are locked at the slower rotation, the phases of the modes vary individually over time at the higher rotation.}
    \label{fig:eigenphaseovertime}
\end{figure}

Figure \ref{fig:eigenphaseovertime} shows the phase of each dominant mode over time, for a case with no oscillation of growth rate and for a case with this oscillation. It can be seen that, in the case with no oscillation, the phases of the mode are different from each other, but change at the same rate and thus maintain the same relative phase over time. For the case with growth rate oscillation, the phases oscillate over time, bringing them further and closer apart in phase. One would expect that the modes will be most in phase with one another at the points in time where maximum growth rate is observed, but further analysis should be performed to check this.

\newpage

\section{Conclusion}

In summary, a radial electric field was introduced to an existing l=2 classical stellarator equilibrium, inducing poloidal flows. The ballooning modes were shown to be stabilized when the shearing rate was comparable to the growth rate in the case without flows, which is consistent with expectations. A unique stabilization curve for all cases was obtained when normalizing appropriately. Oscillations of the linear growth rate of the ballooning modes were observed in cases with higher flows. While the analysis herein points towards several possible causes of this phenomenon, further investigation is required to determine the underlying mechanism of these oscillations. \\

\subsection{Implications and directions for further research}
The poloidal flows were implemented via a radial potential profile, prescribed at the beginning of the simulation and kept constant in the dominant n=0 mode. This was shown to be sufficient for demonstrating the stabilization of the n=10 ballooning mode in the W7-A equilibrium, but may not be sufficient for other cases, for example if the n=0 component interacts significantly with the dynamics of the mode of interest. This method also had the potential to cause numerical instabilities at flow speeds which significantly alter the background equilibrium (see section \ref{sec:prescribingprofile}). Poloidal flows could be modelled for a larger variety of cases by introducing a source term into the time evolution equations, as discussed in section \ref{subsec:alternaltivesolution}. Implementing the source term would thus provide a more robust and widely applicable poloidal flow. This could be used to investigate, for example, shearing rates in more complex geometries, supersonic flow rates, or constant velocity profiles  \cite{wagner_w7-as_2005}.\\

\newpage

\printbibliography

@book{freidberg_plasma_2007,
	address = {Cambridge},
	title = {Plasma physics and fusion energy},
	isbn = {978-0-521-85107-7 978-0-521-73317-5},
	publisher = {Cambridge University Press},
	author = {Freidberg, Jeffrey P.},
	year = {2007},
	note = {OCLC: ocm71808136},
}

@book{noauthor_vmec_nodate,
	title = {{VMEC} wiki},
	url = {https://princetonuniversity.github.io/STELLOPT/},
	urldate = {2023-11-22},
	publisher = {Princeton Plasma Physics Laboratory},
}

@misc{wagner_stellarators_1998,
	title = {Stellarators and {Optimised} {Stellarators}},
	url = {https://www.tandfonline.com/doi/epdf/10.13182/FST98-A11946996?needAccess=true},
	urldate = {2023-10-25},
	author = {Wagner, F},
	year = {1998},
	doi = {10.13182/FST98-A11946996},
}

@article{huysmans_external_2005,
	title = {External kink (peeling) modes in x-point geometry},
	volume = {47},
	issn = {0741-3335, 1361-6587},
	url = {https://iopscience.iop.org/article/10.1088/0741-3335/47/12/003},
	doi = {10.1088/0741-3335/47/12/003},
	language = {en},
	number = {12},
	urldate = {2023-11-07},
	journal = {Plasma Physics and Controlled Fusion},
	author = {Huysmans, G.T.A.},
	month = dec,
	year = {2005},
	pages = {2107--2121},
}

@phdthesis{nikulsin_models_2021,
	type = {Doctoral},
	title = {Models and methods for nonlinear magnetohydrodynamic simulations of stellarators},
	school = {Technische Universität München},
	author = {Nikulsin, Nikita},
	year = {2021},
}

@article{waelbroeck_ballooning_1991,
	title = {Ballooning instabilities in tokamaks with sheared toroidal flows},
	volume = {3},
	issn = {0899-8221},
	url = {https://pubs.aip.org/pfb/article/3/3/601/941306/Ballooning-instabilities-in-tokamaks-with-sheared},
	doi = {10.1063/1.859858},
	language = {en},
	number = {3},
	urldate = {2023-12-07},
	journal = {Physics of Fluids B: Plasma Physics},
	author = {Waelbroeck, F. L. and Chen, L.},
	month = mar,
	year = {1991},
	pages = {601--610},
}

@article{cooper_ballooning_1988,
	title = {Ballooning instabilities in tokamaks with sheared toroidal flows},
	volume = {30},
	issn = {0741-3335, 1361-6587},
	url = {https://iopscience.iop.org/article/10.1088/0741-3335/30/13/001},
	doi = {10.1088/0741-3335/30/13/001},
	number = {13},
	urldate = {2023-12-07},
	journal = {Plasma Physics and Controlled Fusion},
	author = {Cooper, W A},
	month = dec,
	year = {1988},
	pages = {1805--1812},
}

@article{wagner_w7-as_2005,
	title = {W7-{AS}: {One} step of the {Wendelstein} stellarator line},
	volume = {12},
	issn = {1070-664X, 1089-7674},
	shorttitle = {W7-{AS}},
	url = {https://pubs.aip.org/pop/article/12/7/072509/262501/W7-AS-One-step-of-the-Wendelstein-stellarator},
	doi = {10.1063/1.1927100},
	language = {en},
	number = {7},
	urldate = {2023-12-06},
	journal = {Physics of Plasmas},
	author = {Wagner, F. and Bäumel, S. and Baldzuhn, J. and Basse, N. and Brakel, R. and Burhenn, R. and Dinklage, A. and Dorst, D. and Ehmler, H. and Endler, M. and Erckmann, V. and Feng, Y. and Gadelmeier, F. and Geiger, J. and Giannone, L. and Grigull, P. and Hartfuss, H.-J. and Hartmann, D. and Hildebrandt, D. and Hirsch, M. and Holzhauer, E. and Igitkhanov, Y. and Jänicke, R. and Kick, M. and Kislyakov, A. and Kisslinger, J. and Klinger, T. and Klose, S. and Knauer, J. P. and König, R. and Kühner, G. and Laqua, H. P. and Maassberg, H. and McCormick, K. and Niedermeyer, H. and Nührenberg, C. and Pasch, E. and Ramasubramanian, N. and Ruhs, N. and Rust, N. and Sallander, E. and Sardei, F. and Schubert, M. and Speth, E. and Thomsen, H. and Volpe, F. and Weller, A. and Werner, A. and Wobig, H. and Würsching, E. and Zarnstorff, M. and Zoletnik, S.},
	month = jul,
	year = {2005},
	pages = {072509},
}

@article{grieger_wendelstein_1985,
	title = {Wendelstein stellarators},
	volume = {25},
	issn = {0029-5515, 1741-4326},
	url = {https://iopscience.iop.org/article/10.1088/0029-5515/25/9/040},
	doi = {10.1088/0029-5515/25/9/040},
	language = {en},
	number = {9},
	urldate = {2023-11-28},
	journal = {Nuclear Fusion},
	author = {Grieger, G. and Renner, H. and Wobig, H.},
	month = sep,
	year = {1985},
	pages = {1231--1242},
}

@article{hutchinson_single-particle_nodate,
	title = {Single-{Particle} {Motion} in  {Given} {Electric} and {Magnetic} {Fields}},
	language = {en},
	author = {Hutchinson, I H},
}

@article{wilson_non-linear_nodate,
	title = {Non-linear {Ballooning} {Mode} {Theory} and {Consequences} for {ELMs} in {Tokamaks}},
	language = {en},
	author = {Wilson, H R and Cowley, S C and Kirk, A and Snyder, P B},
}

@article{weller_significance_2006,
	title = {Significance of {MHD} {Effects} in {Stellarator} {Confinement}},
	volume = {50},
	issn = {1536-1055, 1943-7641},
	url = {https://www.tandfonline.com/doi/full/10.13182/FST06-A1231},
	doi = {10.13182/FST06-A1231},
	language = {en},
	number = {2},
	urldate = {2023-11-12},
	journal = {Fusion Science and Technology},
	author = {Weller, A. and Sakakibara, S. and Watanabe, K. Y. and Toi, K. and Geiger, J. and Zarnstorff, M. C. and Hudson, S. R. and Reiman, A. and Werner, A. and Nührenberg, C. and Ohdachi, S. and Suzuki, Y. and Yamada, H. and {W7-AS Team} and {LHD Team}},
	month = aug,
	year = {2006},
	pages = {158--170},
}

@article{zhou_approach_2021,
	title = {Approach to nonlinear magnetohydrodynamic simulations in stellarator geometry},
	volume = {61},
	issn = {0029-5515, 1741-4326},
	url = {https://iopscience.iop.org/article/10.1088/1741-4326/ac0b35},
	doi = {10.1088/1741-4326/ac0b35},
	language = {en},
	number = {8},
	urldate = {2023-11-12},
	journal = {Nuclear Fusion},
	author = {Zhou, Yao and Ferraro, N.M. and Jardin, S.C. and Strauss, H.R.},
	month = aug,
	year = {2021},
	pages = {086015},
}

@article{artsimovich_tokamak_1972,
	title = {Tokamak devices},
	volume = {12},
	issn = {0029-5515, 1741-4326},
	url = {https://iopscience.iop.org/article/10.1088/0029-5515/12/2/012},
	doi = {10.1088/0029-5515/12/2/012},
	number = {2},
	urldate = {2023-11-12},
	journal = {Nuclear Fusion},
	author = {Artsimovich, L.A.},
	month = mar,
	year = {1972},
	pages = {215--252},
}

@article{artsimovich_thermal_1967,
	title = {Thermal insulation of plasma in the ?{Tokamaks}?},
	volume = {22},
	issn = {0038-531X, 1573-8205},
	shorttitle = {Thermal insulation of plasma in the ?},
	url = {http://link.springer.com/10.1007/BF01116249},
	doi = {10.1007/BF01116249},
	language = {en},
	number = {4},
	urldate = {2023-11-10},
	journal = {Soviet Atomic Energy},
	author = {Artsimovich, L. A. and Bobrovskii, G. A. and Mirnov, S. V. and Razumova, K. A. and Strelkov, V. S.},
	month = apr,
	year = {1967},
	pages = {325--331},
}

@article{greene_interchange_1968,
	title = {Interchange instabilities in ideal hydromagnetic theory},
	volume = {10},
	issn = {0032-1028},
	url = {https://iopscience.iop.org/article/10.1088/0032-1028/10/8/301},
	doi = {10.1088/0032-1028/10/8/301},
	language = {en},
	number = {8},
	urldate = {2023-11-10},
	journal = {Plasma Physics},
	author = {Greene, John M and Johnson, John L},
	month = jan,
	year = {1968},
	pages = {729--745},
}

@book{braams_nuclear_2002,
	address = {Bristol},
	title = {Nuclear fusion: half a century of magnetic confinement fusion research},
	isbn = {978-0-7503-0705-5},
	shorttitle = {Nuclear fusion},
	language = {en},
	publisher = {Inst. of Physics Publ},
	author = {Braams, Cornelius M. and Stott, Peter E. and Braams, Cornelis M.},
	year = {2002},
}

@article{lehnert_half_2013,
	title = {Half a century of fusion research towards {ITER}},
	volume = {87},
	issn = {0031-8949, 1402-4896},
	url = {https://iopscience.iop.org/article/10.1088/0031-8949/87/01/018201},
	doi = {10.1088/0031-8949/87/01/018201},
	language = {en},
	number = {1},
	urldate = {2023-11-10},
	journal = {Physica Scripta},
	author = {Lehnert, Bo},
	month = jan,
	year = {2013},
	pages = {018201},
}

@article{grieger_physics_1992,
	title = {Physics optimization of stellarators},
	volume = {4},
	issn = {0899-8221},
	url = {https://pubs.aip.org/pfb/article/4/7/2081/940143/Physics-optimization-of-stellarators},
	doi = {10.1063/1.860481},
	language = {en},
	number = {7},
	urldate = {2023-11-07},
	journal = {Physics of Fluids B: Plasma Physics},
	author = {Grieger, G. and Lotz, W. and Merkel, P. and Nührenberg, J. and Sapper, J. and Strumberger, E. and Wobig, H. and Burhenn, R. and Erckmann, V. and Gasparino, U. and Giannone, L. and Hartfuss, H. J. and Jaenicke, R. and Kühner, G. and Ringler, H. and Weller, A. and Wagner, F. and {the W7-X Team} and {the W7-AS Team}},
	month = jul,
	year = {1992},
	pages = {2081--2091},
}

@article{weller_survey_2001,
	title = {Survey of magnetohydrodynamic instabilities in the advanced stellarator {Wendelstein} 7-{AS}},
	volume = {8},
	issn = {1070-664X, 1089-7674},
	url = {https://pubs.aip.org/pop/article/8/3/931/265121/Survey-of-magnetohydrodynamic-instabilities-in-the},
	doi = {10.1063/1.1346633},
	language = {en},
	number = {3},
	urldate = {2023-11-07},
	journal = {Physics of Plasmas},
	author = {Weller, A. and Anton, M. and Geiger, J. and Hirsch, M. and Jaenicke, R. and Werner, A. and {W7-AS Team} and Nührenberg, C. and Sallander, E. and Spong, D. A.},
	month = mar,
	year = {2001},
	pages = {931--956},
}

@article{burrell_effects_1997,
	title = {Effects of {E}×{B} velocity shear and magnetic shear on turbulence and transport in magnetic confinement devices},
	volume = {4},
	issn = {1070-664X, 1089-7674},
	url = {https://pubs.aip.org/pop/article/4/5/1499/980601/Effects-of-E-B-velocity-shear-and-magnetic-shear},
	doi = {10.1063/1.872367},
	language = {en},
	number = {5},
	urldate = {2023-11-01},
	journal = {Physics of Plasmas},
	author = {Burrell, K. H.},
	month = may,
	year = {1997},
	pages = {1499--1518},
}

@article{watanabe_poloidal_1992,
	title = {Poloidal shear flow effect on ideal interchange instability},
	volume = {32},
	issn = {0029-5515},
	url = {https://iopscience.iop.org/article/10.1088/0029-5515/32/9/I13},
	doi = {10.1088/0029-5515/32/9/I13},
	language = {en},
	number = {9},
	urldate = {2023-10-23},
	journal = {Nuclear Fusion},
	author = {Watanabe, K and Sugama, H and Wakatani, M},
	month = sep,
	year = {1992},
	pages = {1647--1652},
}

@article{burrell_role_2020,
	title = {Role of sheared \textit{{E} × {B}} flow in self-organized, improved confinement states in magnetized plasmas},
	volume = {27},
	issn = {1070-664X, 1089-7674},
	url = {https://pubs.aip.org/pop/article/27/6/060501/153617/Role-of-sheared-E-B-flow-in-self-organized},
	doi = {10.1063/1.5142734},
	language = {en},
	number = {6},
	urldate = {2023-10-23},
	journal = {Physics of Plasmas},
	author = {Burrell, K. H.},
	month = jun,
	year = {2020},
	pages = {060501},
}

@article{nikulsin_three-dimensional_2019,
	title = {A three-dimensional reduced {MHD} model consistent with full {MHD}},
	volume = {26},
	issn = {1070-664X, 1089-7674},
	url = {http://arxiv.org/abs/1907.12486},
	doi = {10.1063/1.5122013},
	language = {en},
	number = {10},
	urldate = {2023-09-26},
	journal = {Physics of Plasmas},
	author = {Nikulsin, Nikita and Hoelzl, Matthias and Zocco, Alessandro and Lackner, Karl and Günter, Sibylle},
	month = oct,
	year = {2019},
	note = {arXiv:1907.12486 [physics]},
	pages = {102109},
}

@article{nikulsin_jorek3d_2022,
	title = {{JOREK3D}: {An} extension of the {JOREK} nonlinear {MHD} code to stellarators},
	volume = {29},
	issn = {1070-664X, 1089-7674},
	shorttitle = {{JOREK3D}},
	url = {https://pubs.aip.org/pop/article/29/6/063901/2848122/JOREK3D-An-extension-of-the-JOREK-nonlinear-MHD},
	doi = {10.1063/5.0087104},
	language = {en},
	number = {6},
	urldate = {2023-09-22},
	journal = {Physics of Plasmas},
	author = {Nikulsin, N. and Ramasamy, R. and Hoelzl, M. and Hindenlang, F. and Strumberger, E. and Lackner, K. and Günter, S. and {JOREK Team}},
	month = jun,
	year = {2022},
	pages = {063901},
}

@article{hameiri_equilibrium_1983,
	title = {The equilibrium and stability of rotating plasmas},
	volume = {26},
	issn = {0031-9171},
	url = {https://pubs.aip.org/pfl/article/26/1/230/815814/The-equilibrium-and-stability-of-rotating-plasmas},
	doi = {10.1063/1.864012},
	language = {en},
	number = {1},
	urldate = {2023-09-11},
	journal = {The Physics of Fluids},
	author = {Hameiri, Eliezer},
	month = jan,
	year = {1983},
	pages = {230--237},
}

@article{nuhrenberg_quasi-helically_1988,
	title = {Quasi-helically symmetric toroidal stellarators},
	volume = {129},
	issn = {03759601},
	url = {https://linkinghub.elsevier.com/retrieve/pii/0375960188900801},
	doi = {10.1016/0375-9601(88)90080-1},
	language = {en},
	number = {2},
	urldate = {2023-05-23},
	journal = {Physics Letters A},
	author = {Nührenberg, J. and Zille, R.},
	month = may,
	year = {1988},
	pages = {113--117},
}

@article{hoelzl_jorek_2021,
	title = {The {JOREK} non-linear extended {MHD} code and applications to large-scale instabilities and their control in magnetically confined fusion plasmas},
	volume = {61},
	issn = {0029-5515, 1741-4326},
	url = {https://iopscience.iop.org/article/10.1088/1741-4326/abf99f},
	doi = {10.1088/1741-4326/abf99f},
	language = {en},
	number = {6},
	urldate = {2023-05-10},
	journal = {Nuclear Fusion},
	author = {Hoelzl, M. and Huijsmans, G.T.A. and Pamela, S.J.P. and Bécoulet, M. and Nardon, E. and Artola, F.J. and Nkonga, B. and Atanasiu, C.V. and Bandaru, V. and Bhole, A. and Bonfiglio, D. and Cathey, A. and Czarny, O. and Dvornova, A. and Fehér, T. and Fil, A. and Franck, E. and Futatani, S. and Gruca, M. and Guillard, H. and Haverkort, J.W. and Holod, I. and Hu, D. and Kim, S.K. and Korving, S.Q. and Kos, L. and Krebs, I. and Kripner, L. and Latu, G. and Liu, F. and Merkel, P. and Meshcheriakov, D. and Mitterauer, V. and Mochalskyy, S. and Morales, J.A. and Nies, R. and Nikulsin, N. and Orain, F. and Pratt, J. and Ramasamy, R. and Ramet, P. and Reux, C. and Särkimäki, K. and Schwarz, N. and Singh Verma, P. and Smith, S.F. and Sommariva, C. and Strumberger, E. and Van Vugt, D.C. and Verbeek, M. and Westerhof, E. and Wieschollek, F. and Zielinski, J.},
	month = jun,
	year = {2021},
	pages = {065001},
}

@article{boozer_stellarator_2015,
	title = {Stellarator design},
	volume = {81},
	issn = {0022-3778, 1469-7807},
	url = {https://www.cambridge.org/core/product/identifier/S0022377815001373/type/journal_article},
	doi = {10.1017/S0022377815001373},
	language = {en},
	number = {6},
	urldate = {2023-05-06},
	journal = {Journal of Plasma Physics},
	author = {Boozer, Allen H.},
	month = dec,
	year = {2015},
	pages = {515810606},
}

@techreport{j_d_lawson_criteria_1955,
	address = {Harwell, Berkshire, U. K.},
	title = {Some {Criteria} for a {Useful} {Thermonuclear} {Reactor}},
	url = {https://www.euro-fusion.org/fileadmin/user_upload/Archive/wp-content/uploads/2012/10/dec05-aere-gpr1807.pdf},
	institution = {Atomic Energy Research Establishment},
	author = {{J. D. Lawson}},
	month = dec,
	year = {1955},
}

@article{ongena_magnetic-confinement_2016,
	title = {Magnetic-confinement fusion},
	volume = {12},
	issn = {1745-2473, 1745-2481},
	url = {http://www.nature.com/articles/nphys3745},
	doi = {10.1038/nphys3745},
	language = {en},
	number = {5},
	urldate = {2022-03-06},
	journal = {Nature Physics},
	author = {Ongena, J. and Koch, R. and Wolf, R. and Zohm, H.},
	month = may,
	year = {2016},
	pages = {398--410},
}

@incollection{zohuri_confinement_2017,
	address = {Cham},
	title = {Confinement {Systems} for {Controlled} {Thermonuclear} {Fusion}},
	isbn = {978-3-319-51176-4 978-3-319-51177-1},
	url = {http://link.springer.com/10.1007/978-3-319-51177-1_3},
	language = {en},
	urldate = {2022-03-06},
	booktitle = {Magnetic {Confinement} {Fusion} {Driven} {Thermonuclear} {Energy}},
	publisher = {Springer International Publishing},
	author = {Zohuri, Bahman},
	collaborator = {Zohuri, Bahman},
	year = {2017},
	doi = {10.1007/978-3-319-51177-1_3},
	pages = {103--182},
}

@article{azarpour_review_2013,
	title = {A {Review} on the {Drawbacks} of {Renewable} {Energy} as a {Promising} {Energy} {Source} of the {Future}},
	volume = {38},
	issn = {1319-8025, 2191-4281},
	url = {http://link.springer.com/10.1007/s13369-012-0436-6},
	doi = {10.1007/s13369-012-0436-6},
	language = {en},
	number = {2},
	urldate = {2022-02-28},
	journal = {Arabian Journal for Science and Engineering},
	author = {Azarpour, Abbas and Suhaimi, Suardi and Zahedi, Gholamreza and Bahadori, Alireza},
	month = feb,
	year = {2013},
	pages = {317--328},
}

@article{helander_stellarator_2012,
	title = {Stellarator and tokamak plasmas: a comparison},
	volume = {54},
	issn = {0741-3335, 1361-6587},
	shorttitle = {Stellarator and tokamak plasmas},
	url = {https://iopscience.iop.org/article/10.1088/0741-3335/54/12/124009},
	doi = {10.1088/0741-3335/54/12/124009},
	number = {12},
	urldate = {2022-02-27},
	journal = {Plasma Physics and Controlled Fusion},
	author = {Helander, P and Beidler, C D and Bird, T M and Drevlak, M and Feng, Y and Hatzky, R and Jenko, F and Kleiber, R and Proll, J H E and Turkin, Yu and Xanthopoulos, P},
	month = dec,
	year = {2012},
	pages = {124009},
}

@article{xu_general_2016,
	title = {A general comparison between tokamak and stellarator plasmas},
	volume = {1},
	issn = {2468-2047},
	url = {https://aip.scitation.org/doi/10.1016/j.mre.2016.07.001},
	doi = {10.1016/j.mre.2016.07.001},
	number = {4},
	urldate = {2022-02-27},
	journal = {Matter and Radiation at Extremes},
	author = {Xu, Yuhong},
	month = jul,
	year = {2016},
	pages = {192--200},
}

@book{wesson_tokamaks_2004,
	address = {Oxford : New York},
	edition = {3rd ed},
	series = {Oxford science publications},
	title = {Tokamaks},
	isbn = {978-0-19-850922-6},
	number = {118},
	publisher = {Clarendon Press ; Oxford University Press},
	author = {Wesson, John and Campbell, D. J.},
	year = {2004},
}

\end{document}